\documentclass[12pt,preprint]{aastex}
\input epsf
\usepackage{graphicx,natbib}
\def \half {{\case{1}{2}}}

\begin{document}

\title{Optical Cluster-Finding with An Adaptive Matched-Filter Technique: Algorithm 
and Comparison with Simulations }
\author{Feng Dong \altaffilmark{1}, Elena Pierpaoli \altaffilmark{2}, 
James E. Gunn \altaffilmark{3}, Risa H. Wechsler \altaffilmark{4}}
\altaffiltext{1}{Department of Astrophysical Sciences, Princeton University, Princeton, 
NJ 08544, feng@astro.princeton.edu}
\altaffiltext{2}{University of Southern California, Los Angels, CA, 90089-0484, pierpaol@usc.edu}
\altaffiltext{3}{Department of Astrophysical Sciences, Princeton University, Princeton, 
NJ 08544, jeg@astro.princeton.edu}
\altaffiltext{4}{Kavli Institute for Particle Astrophysics and Cosmology, 
Physics Department, and
Stanford Linear Accelerator Center, 
Stanford University,
Stanford, CA 94305}

\begin{abstract}

We present a modified adaptive matched filter algorithm designed to identify clusters of galaxies
in wide-field imaging surveys such as the Sloan Digital Sky Survey. The cluster-finding technique 
is fully adaptive to imaging surveys with spectroscopic coverage, multicolor photometric redshifts, 
no redshift information at all, and any combination of these within one survey. It works with high 
efficiency in multi-band imaging surveys where photometric redshifts can be estimated with well-understood 
error distributions. Tests of the algorithm on realistic mock SDSS catalogs suggest that the detected 
sample is $\sim 85\%$ complete and over $90\%$ pure for clusters with masses above $1.0\times10^{14} h^{-1}$ 
M$_\odot$ and redshifts up to $z=0.45$. The errors of estimated cluster redshifts from maximum likelihood 
method are shown to be small (typically less that 0.01) over the whole redshift range with photometric 
redshift errors typical of those found in the Sloan survey. Inside the spherical radius corresponding 
to a galaxy overdensity of $\Delta=200$, we find the derived cluster richness $\Lambda_{200}$ a roughly 
linear indicator of its virial mass $M_{200}$, which well recovers the relation between total 
luminosity and cluster mass of the input simulation.
\end{abstract}


\section{Introduction} \label{sec:Introduction}

Clusters of galaxies are the most massive virialized systems in the Universe and have been 
extensively used to study galaxy population and evolution \citep{Dre84, Dre92}, to trace 
the large-scale structure of the universe \citep{Bah88, Pos92}, and to constrain cosmology 
\citep{Evr89, Bah99, Hen00,Pierpa01,Pierpa03}. Given the important roles clusters of galaxies 
play in the studies of both astrophysics and cosmology, tremendous efforts have been made 
during the past several decades to search for these systems. The first large samples of 
clusters were identified by looking for projected galaxy overdensities through visual 
inspection of photographic plates \citep{Abe58, Abe89, Zwi68}. These catalogs made pioneering 
contributions to our understanding of the extragalactic universe and since their generation 
have opened many new frontiers in the studies of galaxy clusters. However, the compilation 
of a relatively complete and pure sample of galaxy clusters has remained far from trivial. 
To date the Abell catalog, which contains about 4000 rich clusters to a redshift of $z\sim0.2$, 
is still the most widely used cluster catalog in the field, though it was realized early that 
visually-constructed catalogs suffer from projection effects, subjectivity, and large 
uncertainties in estimated properties \citep{Sut88}. It is difficult to apply these catalogs 
for statistical studies in cosmology because of these uncertainties, in addition to the fact 
that the selection function and false positive rates of such cluster samples are hard to quantify.

To relieve some of these concerns, other approaches for identifying clusters have also been 
designed and implemented, such as reconstructing the full 3-D structures in complete redshift 
surveys \citep{Huc82, Gel83, Ram97}, detecting clusters in X-ray surveys \citep{Gio90, Edg90, 
Ebe98, Ros98, Rom00, Sch00, Boh01, Mul03, Boh04}, and utilizing the Sunyaev-Zeldovich effect 
\citep{Car00, Moh02, Pierpa05} and weak gravitational lensing \citep{Sch96, Wit01} in search for 
clusters.  Moreover, the realization of large and deep galaxy surveys in recent years has 
revived optical cluster-finding endeavors and prompted the development of more automated 
and rigorous algorithms to select clusters from imaging surveys. Using multi-color photometric 
data from which photometric redshifts can be estimated, it is now possible to mitigate the 
problems of projection effects, and quantitative analysis of the selection bias is also now 
possible. Automated peak-finding techniques in optical cluster searches were attempted by 
\citet{She85} and later used in the Edinburgh/Durham survey \citep[ED][]{Lum92} as well as the 
Automatic Plate Measurement Facility survey \citep[APM][]{Dal94, Dal97}. In the construction 
of the cluster catalog from the Palomar Distant Cluster Survey \citep{Pos96}, a matched filter 
algorithm was developed to select clusters from a photometric galaxy sample. It was widely used 
in subsequent surveys and several variants have been put forward \citep{Kaw98, Sch98, Kep99, Kim02, Whi02}. 
Meanwhile with the knowledge of the existence of the ``E/S0 ridgeline'' of cluster galaxies in 
color-magnitude space and the aid of multi-color CCD photometry, several color-based cluster-finding 
techniques were also investigated \citep{Gla00, Got02, Gla05, Mil05}. Some of these have already 
been successfully applied to select clusters from the Sloan Digital Sky Survey (SDSS) data 
\citep{Got02, Ann02, Bah03, Mil05, Koe07}.

The Sloan Digital Sky Survey \citep{Yor00} is a five-band CCD imaging survey of about 10$^4$ deg$^{2}$ 
in the high latitude North Galactic Cap and a smaller deeper region in the South, followed by 
an extensive multi-fiber spectroscopic survey. The imaging survey is carried out in drift-scan 
mode in five SDSS filters ($u$, $g$, $r$, $i$, $z$) to a limiting magnitude of $r\sim22.5$ 
\citep{Fuk96, Gun98, Lup01, Smi02}. The spectroscopic survey targets $\sim$10$^{6}$ galaxies 
to  $r\sim17.7$, with a median redshift of $z\sim0.1$ \citep{Str02}, and a smaller deeper sample 
of $\sim$10$^{5}$ Luminous Red Galaxies out to $z\sim0.5$ \citep{Eis01}. In this paper we discuss 
a modified adaptive matched filter technique incorporating several new features over previous 
algorithms and designed to detect clusters using both the SDSS imaging and spectroscopic data; 
it could readily be adapted to other similar multi-band, large-area galaxy surveys for construction 
of optically-selected cluster samples. It is the first of a series of papers that will explore 
the application of the technique to select clusters from the Sloan Digital Sky Survey.

The general idea of the matched filter method relies on the fact that clusters show on average 
a typical density profile, now widely assumed to be the ``NFW'' form suggested first by Navarro, 
Frenk and White \citep{Nav96}. Assuming that galaxies trace the dark matter, we expect galaxies 
within clusters to be distributed according to such profile. The algorithm selects regions in 
the sky where the distribution of galaxies corresponds to the projection of average cluster 
density profile. In addition, it is possible to specify the galaxy redshift information 
inside clusters, and to use prior knowledge on the galaxy luminosity function. The combination 
of these matched subfilters thus enables us to extract a quantitative signal corresponding to 
the existence of a cluster at a given location in the surveyed sky area.

The modified matched filter technique presented in this paper can fully adapt to imaging surveys 
with spectroscopic measurements, multicolor photometric redshifts, no redshift information at all, 
and any combination of these within one survey. In the Sloan Digital Sky Survey where photometric 
redshifts can be estimated with well-understood error distributions from the five-band 
($u$,$g$,$r$,$i$,$z$) multi-color photometry, the matched filter technique described here utilizes 
not only the spectroscopic coverage for the bright main sample galaxies and Luminous Red Galaxies (LRGs) 
but also the photometric redshift information for most of the galaxies detected in the imaging 
survey. This greatly expands the input galaxy sample to feed into the cluster-finding algorithm 
compared to pure spectroscopic methods \citep[e.g.][]{Mil05}. The obtained composite cluster catalog 
can also go much deeper in redshift ($z\sim0.4-0.5$ in this case) than the typical $z\sim0.2$ limit 
for spectroscopic samples due to the lack of availability of spectroscopic measurements for faint, 
deep galaxies.

Since the matched filter technique does not explicitly use the information about the red sequence to 
select clusters as is done in some color-based cluster-finding methods \citep{Ann02, Mil05, Koe07}, 
it can theoretically detect clusters of any type in color, and is not restricted only to old, red 
E/S0 galaxies. Such clusters likely dominate the cluster population, but may not constitute all 
of it especially as one probes systems of lower richness and at higher redshifts. The use of both 
spectroscopic and photometric redshift information largely eliminates the projection effects and 
removes most of the phantom clusters. The matched filter also generates accurate quantitative estimates 
of derived cluster properties, such as redshift, scale, richness, and concentration, and produces 
quantitative detection likelihoods, indicative of the combined information for both red and blue 
galaxies identified as cluster members. These facilitate further studies of detected systems and 
makes easier the comparison to clusters selected by other methods. One major concern for the matched 
filter technique is the fact that determination of these parameters depends on the specific cluster 
model we put in to build the relevant filters. However, these effects can be minimized by careful 
assumptions about the shape and evolution of luminosity function, and by the fact that our density 
filter is self-adaptive to different cluster scales and concentration. The clusters selected by 
the algorithm will provide us the necessary sample on which we then apply an iterative procedure 
aimed at refining the constraints on clusters' properties. More details will be discussed in 
section \S\ref{sec:Algorithm} and subsequent work following this paper.

The new algorithm presented here differs from previous matched filter implementations 
\citep{Kep99, Kim02} in several ways. We use a uniform Poisson likelihood analysis, which is only 
the second step in the approach by \citet{Kep99} following a first pass using Gaussian statistics for 
pre-selection of clusters. This avoids the common problem for high-redshift clusters of having 
too few galaxies in any cell of interest for Gaussian statistics to apply, and the adopted approach 
yields correct likelihoods even at the detection stage. In addition, both the core radius and 
virial radius of the matched filter are adaptive over the typical observed dynamical range for clusters,
in contrast to most previous cluster-finding techniques that set the cluster core radius or search 
radius to be fixed. For each individual cluster, a best-fit core radius is found to maximize the 
likelihood match, as well as an outer radius inside which the galaxy overdensity reaches $\Delta$=200. 
The cluster richness is then normalized to be the light contained within this virial radius, which we 
find correlates better with the mass of gravitational systems whose extent is defined by density 
contrast as is widely adopted in theoretical studies. The new features of our modified algorithm 
will be further discussed in \S\ref{sec:Algorithm}.

In order to understand the biases and the selection functions of our algorithm, we test it on 
a mock SDSS catalog which has been constructed from the Hubble Volume Simulation \citep{Evr02} 
by assigning luminosities and colors to the dark matter particles in a manner which reproduces 
many characteristics of the galaxy population from SDSS observations. The ``observations'' of 
the simulations have then been further modified so that the redshift scatter of those galaxies 
which have photometric but no spectroscopic redshifts correspond to that of the photometric redshift 
errors in actual SDSS data. The comparison of the detected cluster sample with halos in 
the simulation provides the only rigorous way to assess how the observed cluster properties 
relate to the real masses, and how the cluster sample can be used to derive cosmological 
constraints.

In section \S\ref{sec:Algorithm} we describe the modified adaptive matched filter technique 
and how it is used to extract the cluster sample. Section \S\ref{sec:Simulation} presents 
the basic features of the simulated catalog we adopted for the testing purpose. In section 
\S\ref{sec:Results} we show results on the completeness and purity of our cluster sample, and 
the expected scaling relations inferred from runs on the simulations. We conclude in section 
\S\ref{sec:Conclusion}.

A flat $\Lambda$CDM model with $\Omega_m=0.3$ and $\Omega_{\Lambda}=0.7$ is used throughout 
this work, and we assume a Hubble constant of $H_0=100 h$ km s$^{-1}$ Mpc$^{-1}$ if not 
specified otherwise.

\section{The Cluster-Finding Algorithm} \label{sec:Algorithm}

The matched filter technique introduced here is a likelihood method which identifies clusters 
by convolving the optical galaxy survey with a set of filters based on a modeling of the cluster 
and field galaxy distributions. A cluster radial surface density profile, a galaxy luminosity 
function, and redshift information (when available) are used to construct filters in position, 
magnitude, and redshift space, from which a cluster likelihood map is generated. The peaks in 
the map thus correspond to candidate cluster centers where the matches between the survey data 
and the cluster filters are optimized. The algorithm automatically provides the probability for 
the detection, best-fit estimates of cluster properties including redshift, radius and richness, 
as well as membership assessment for each galaxy. The modified algorithm can be fully adaptive 
to current and future galaxy surveys in 2-D (imaging), 2$\half$-D (where multi-color photometric 
redshifts and their errors can be estimated), and 3-D (with full spectroscopic redshift measurements). 
Usage of the apparent magnitudes and, where applicable, the redshift estimates instead of simply 
searching for projected galaxy overdensities effectively suppresses the foreground-background 
contamination, and the technique has proven to be an efficient way of selecting clusters of 
galaxies from large multi-band optical surveys.

In what follows, we first provide a general introduction on how the likelihood function is 
constructed and how we detect clusters with the matched filter method. This gives us an overview 
about how the cluster catalog is derived. Then we discuss in more detail the density models and 
subfilters used to construct the likelihood. More specifically, we assume an NFW density profile, 
a general Schechter luminosity function and a Gaussian model for BCGs to model clusters, and use 
the spectroscopic measurements and obtained error distributions of galaxy photometric redshifts 
from the Sloan Digital Sky Survey to incorporate redshift uncertainties. In the end we describe how 
to determine the set of best-fit parameters on cluster properties that maximize the likelihood at 
a given position over a range of redshift, scale, concentration, and richness.

\subsection{Likelihood Function} \label{sec:Likelihood}

The likelihood function used here is based on the assumption that the probability of finding galaxies 
in an infinitesimal bin in angular position, apparent magnitude and redshift space is given by a Poisson 
distribution. Under this assumption, the total likelihood of many of such bins, which we take to be 
centered in the location of the galaxies in the survey, is (see appendix C2 in \citet{Kep99} for a full 
derivation):

\begin{equation}
ln {\cal L} = -N_f -\sum_{k=1}^{N_c}N_k +\sum_{i=1}^{N_g}\ln[P(i)], 
\end{equation}

where $N_f$ is the total number of field galaxies expected within the searching area, $N_g$ is the total 
number of galaxies and $\sum_{k=1}^{N_c}N_k$ is normalized to be the number of galaxies 
brighter than $L^*$ as members of the $N_c$ clusters assumed in the model. $P(i)$ represents the 
predicted probability density of galaxies in a given bin, which includes both probabilities 
of field galaxies ($P_f$) and of cluster members ($P_c$),

\begin{equation}
  P(i) = P_f(i) + \sum_{k=1}^{N_c} P_{c}(i,k).
\end{equation}

These probabilities are the expected number densities for a given location and magnitude.

The cluster catalog is constructed with an iterative procedure similar to the one used in \citet{Koc03}.
We start our process from a density model of a smooth background with no clusters. For each galaxy 
position, we then evaluate the likelihood increment we would obtain by assuming that there is in fact 
a cluster centered on that galaxy. The likelihood is then optimized by varying the cluster galaxy 
number $N_k$, the redshift and cluster scale length. At each iteration, we retain the cluster candidate 
which resulted in the greatest likelihood increase. We incorporate it in our density model and restart 
the procedure. The function for finding the $k^{th}$ cluster in the whole surveyed area therefore is 

\begin{equation}
\begin{array}{ll}
  \Delta\ln{\cal L}(k) = -N_k + \sum_{i=1}^{N_g}\ln[{P_f(i)+\sum_{j=1}^{k}P_c(i,j) \over 
      P_f(i)+\sum_{j=1}^{k-1}P_c(i,j)}].
\end{array}
\end{equation}

A list of cluster candidates then becomes available in decreasing order of detection likelihoods.
For each candidate one has derived properties, including best-fit position, scale, richness, and 
estimated redshift. The initial cluster catalog allows us to further inspect each individual candidate 
for exploration of substructure and better constraints on previously fitted quantities.

\subsection{Density Model} \label{sec:Density}

As both field and cluster galaxies are found in the survey, the probability of finding a galaxy in 
a given bin depends on the density of both these populations (see eq.(2)).

For galaxy $i$ with angular position $\vec{\theta}_i$, $r$-band apparent magnitude $m^r_i$ and 
redshift $z_i$ (when available), the background number density $P_f(i)$ can be directly extracted 
from the global number counts of the galaxy survey,  

\begin{equation}
  P_f(i) = { d N \over dm~ dz }(m^r_i, z_i),  
\end{equation}

and it has to be modified to account for the effects of galaxy redshift uncertainties if photometric 
redshift estimates are used.

For cluster $k$ located at $\vec{\theta}_k$ with proper scale length $r_{ck}$, redshift $z_k$ and 
galaxy number $N_k$, the probability of galaxy $i$ being a member of it, $P_c(i,k)$, is just the 
product of a surface density profile $\Sigma_c$ and a luminosity function $\phi_c$ at the cluster's 
redshift, times a distribution function $f(z_i- z_k)$ that expresses redshift uncertainties: 

\begin{equation}
  P_c(i,k) = N_{k}\ \Sigma_c\left[D_A(z_k)\theta_{ik}\right]\ \phi_c\left[m^r_i-{\cal D}(z_k);
    \right]\ f(z_i-z_k),
\end{equation}

where ${\cal D}(z_k)$ is defined through 

\begin{equation}
  M^r_i = m^r_i - 5 \log (D_L(z_k)/10\mbox{pc}) - k(z_k) = m^r_i - {\cal D}(z_k),
\end{equation}

and where $D_A(z_k)$ and $D_L(z_k)$ are the angular diameter and luminosity distance at the cluster's 
redshift $z_k$,  and $k(z_k)$ is the $k$-correction. The conversion of units in luminosity and 
distance is conducted by performing proper $k$-corrections for galaxies of different spectral types 
and choosing the proper cosmology (see \S\ref{sec:Introduction}).

\subsection{Subfilters} \label{sec:Subfilters}

Based on current observational studies as well as findings from dark matter halos, and for convenient 
comparisons to theoretical models widely used in analytical studies and N-body simulations, we assume 
the density profile of galaxies within a cluster follows the form of a NFW profile \citep{Nav96}, 
which in three dimensions is given by 

\begin{equation}
  \rho_c(r) = {1 \over 4\pi r_c^3 F(c)}{1 \over {r \over r_c}(1+{r \over r_c})^2},
\end{equation}

where $c$ is the concentration parameter and $F(c)$ is the typical normalization factor for galaxies 
inside the virial radius of the cluster, $r_v=cr_c$. The 3-D profile is then integrated along the 
line of sight to derive a projected surface density profile $\Sigma_c(r)$ which is expressible as 
a much more complicated analytical form (see \citealt{Bar96}). The profile is normalized so that 
$\int_0^{cr_c} 2\pi r \Sigma_c(r) dr = 1$.

The search radius for galaxies belonging to the cluster is set to be the virial radius of the 
cluster, or more specifically here, the radius inside which the mass overdensity is 200 times 
the critical density, i.e., 200$\Omega_M^{-1}$ times the average background \citep{Evr02}.
Since it is hard to directly measure the cluster mass overdensity in observations, we instead 
determine the virial radius inside which the space density of cluster galaxies is 200$\Omega_M^{-1}$ 
times the mean field, assuming that the galaxy distribution in a halo traces the overall dark 
matter distribution (see discussions in \citealt{Han05}), which has been suggested by recent 
observations and simulations \citep{Lin04a, Nag05, Lin07a}, and is supported by weak lensing 
measurements \citep{She04}. For simplicity, we use $r_{200}$ throughout this work to denote 
the cluster virial radius determined by galaxy overdensities. The cluster richness is then 
defined to be the total luminosity in units of $L^*$ inside $r_{200}$.

As has been discussed before in matched-filter studies \citep{Pos96, Kim02} and also shown by 
our own numerical experiments, the efficiency of the filter is usually much more sensitive to 
the overall filter cutoff radius than to the details of its shape. Therefore the determination 
of appropriate values for the scale length in the cluster model is of particular importance, 
as it may have significant impact on the detection efficiency of the cluster-finding algorithm. 
Most of the previous matched filter methods have used a carefully chosen fixed value for the 
model cluster cutoff radii, and they compute the galaxy number or the richness of clusters within 
such a fixed radius in physical units. \citet{Pos96} concludes that a fixed search radius of 
1 Mpc $h^{-1}$ is a near-optimal choice in their radial filter, and this value has been also 
adopted by \citet{Kep99, Kim02} in their method which assumes a modified Plummer law model 
for the surface density profile. In \citet{Whi02} and \citet{Koc03}, the authors set a fixed 
core radius of $r_c=200$ kpc $h^{-1}$ and concentration parameter of $c=4$ for the NFW profile 
in the cluster detection and mass estimates. Although we find from observations and simulations 
that these choices are reasonable values for typical rich clusters, a single fixed scale length 
for all clusters over a wide range of masses and concentrations will certainly degrade the 
signal-to-noise ratio, bias detection probabilities, and be responsible for at least part of 
the large scatter observed in previous cluster mass-richness scaling relations. In our modified 
adaptive matched filter algorithm, we optimize the core radius for each individual cluster over 
the dynamical range for typical galaxy clusters. For the core radius value that maximizes the 
likelihood, we then compute the normalized cluster richness according to the NFW profile with 
best-fit parameters within a cluster virial radius $r_{200}$ determined from galaxy overdensities. 
We believe this procedure is more similar to and comparable with the virial mass defined by 
density contrast in most theoretical studies and analyses of simulations.

For the magnitude filter, we adopt a luminosity profile described by a central galaxy plus
a standard Schechter luminosity function \citep{Sch76}

\begin{equation}
\phi(M) = {dn \over dM} = 0.4 \ln10\ n^* \left({L \over L^*} \right)^{1+\alpha} \exp(-L/L^*);
\end{equation}

the integrated luminosity function is 

\begin{equation}
\Phi(M) = \int_{-\infty}^M \phi(M) dM = n^* \Gamma[1+\alpha, L/L^*].
\end{equation}

Parameters for the global luminosity function are obtained from the SDSS spectroscopic sample at 
the redshift of $z=0.1$ \citep{Bla03}. To account for the evolutionary effects at higher redshifts, 
we allow a passive evolution of $L^*$ which brightens about 0.8 magnitudes from $z=0$ to $z=0.5$ 
\citep{Lov92, Lil95b, Nag01, Bla03, Lov04, Bal05, Ilb05}. We assume that $L^*$ does not vary as a 
function of cluster richness, which is supported by the results of \citet{Han05}. Because the 
matched filter algorithm uses both a cluster galaxy luminosity function and a field galaxy 
luminosity function, which are expected to be different due to the morphology-density relation 
\citep{Dre80} and the observed dependence of luminosity function on galaxy over-densities 
\citep{Chr00, Mo04, Cro05}, it would be desirable to model these separately. It would also be 
desirable to further model the luminosity distributions according to galaxy spectral types 
\citep{Fol99, Lin99, Hog03}. At this stage, however, only a single function is adopted since 
the work on precise luminosity functions for cluster galaxies of different types has just been 
started. We hope to investigate this further on the basis of the first catalog we produce. 
Once a cluster catalog is available for galaxies in all redshift ranges, we can go back and 
examine the impact of our assumptions about the galaxy luminosity functions as well as their 
evolution for different environments and spectral types. In order to use the same range in 
the luminosity function at all distances and therefore avoid bias associated with errors in the 
assumed form of the luminosity function, we cut off the luminosity function at one magnitude 
below $L^*$.  We can still calculate total luminosities by integrating the assumed form, and 
we use this in our richness calculation, described below.

The existence of Brightest Cluster Galaxies (BCGs) near the cluster centers is incorporated into 
our cluster galaxy luminosity model as a separate component from the main Schechter function for 
satellites, as this distinction has been clearly seen in clusters over a range of richness \citep{RT77,Han05}. 
We assume a Gaussian distribution for the luminosities of these objects and adopt the results 
from \citet{Lin04b} for correlations between the BCG luminosity and host cluster properties. 
More specifically, the BCG luminosity is assumed to follow a single power law with the cluster 
richness, $L_{BCG} \sim \Lambda_{200}^{1\over4}$, and we take the width of the Gaussian to be 
$\sim0.5$ mag \citep{Lin04b, Zhe05, Han05}. The luminosity of BCGs is assumed to evolve in the 
same way as $L^*$ does, {\it i.e.~} the luminosity at the mean of the gaussian has a constant 
ratio to $L^*$. This is almost certainly incorrect in detail, but will be explored in follow-up 
work once the catalog is constructed. This modification of the 
general Schechter function enhances the detectability of typical clusters with BCGs, especially 
those at higher redshifts with only few galaxies other than the BCG to be included in the 
apparent magnitude-limited galaxy sample.

Thanks to the accurate five-band ($u$,$g$,$r$,$i$,$z$) multi-color photometry in the SDSS 
\citep{Yor00}, as well as the associated redshift survey for the bright main sample galaxies 
\citep{Str02} and Luminous Red Galaxies (LRGs, \citealt{Eis01}), it is now also possible to retrieve 
redshift information for most of the galaxies that we are going to use in construction of the 
SDSS cluster catalog, either photometrically or spectroscopically. For real SDSS data 
currently available from DR5, we find that galaxies with valid photometric redshift estimates make 
up more than $96\%$ of the whole sample in the imaging data, within which about $1\%$, mostly 
bright, red galaxies, have matched spectroscopic measurements from redshift surveys. 
Not surprisingly, the inclusion of galaxy redshift estimates greatly improves the accuracy of 
the cluster redshift determinations and significantly mitigates projection effects, thus allowing 
the detection of much poorer systems than possible in previous work with no redshift measurements.

The uncertainties of galaxy redshifts are assumed to follow Gaussian distributions in the 2$\half$-D 
and 3-D cases, where in terms of the $f(z)$ function in equation (5) we have     

\begin{equation}
  f(z_k) = {\exp\left[-(z_i-z_k)^2/2\sigma_{z_i}^2\right] \over \sqrt{2\pi}\sigma_{z_i}}.   
\end{equation}

For galaxies with computed photometric redshifts (described below), we add to the cluster galaxy 
density model a third subfilter based on the distribution of derived redshift uncertainties in the 
form of a combination of multiple Gaussian modes. These error estimates are obtained by calibrating 
photometric redshifts with the real redshifts in the SDSS spectroscopic galaxy sample and redshifts 
for other fainter (but smaller) overlapping surveys. The analysis is done for red and blue galaxies 
separately using the color separator by \citet{Strv01}, and it is found that a model using Gaussian 
modes with proper weights assigned generally provides a good description of the bias and scatter 
in the photometric redshifts for galaxies of both spectral types and in different apparent magnitude 
bins. Some of the results are shown in \S\ref{sec:Simulation}.

In the 3-D case where spectroscopic redshifts of galaxies are measured, we smooth them in 
Gaussians with assigned cluster velocity dispersions that vary in the range from $400$ km s$^{-1}$ 
(proper) for poorer clusters to $1200$ km s$^{-1}$ (proper) for the richest systems in the 
selected cluster sample, according to several discrete estimated richness classes. The same 
procedure as outlined in the previous paragraph for photometric redshifts is applied to include 
this redshift filter in the galaxy density model.

In addition, there are galaxies we find that either have invalid photometric redshifts computed 
or fall into the redshift and magnitude range where no good calibrations are available. Such 
galaxies, which are currently about $3\%$-$5\%$ of the whole sample, are assumed to have no redshift 
estimates and therefore no constraining filter. Hence we set up for each galaxy the appropriate 
scenario that adapts the matched filter algorithm to galaxy redshift estimates with varied accuracy.

Finally, of course, we fit an overall amplitude, which represents the cluster richness. Since
its size, shape and redshift are all determined at this point, we can express the amplitude
however we like in physical terms. We have chosen to use the total luminosity within $r_{200}$
expressed as a multiple of $L^*$ (evolved to the relevant redshift using 1.6 mags of luminosity
evolution per unit redshift), which we denote as $\Lambda_{200}$.

\subsection{Implementation} \label{sec:Implementation}

Implementation of the matched filter algorithm starts with reading the galaxy catalog. For each 
galaxy $i$ in the sample, we read in the positions $\alpha_i$, $\delta_i$, the extinction-corrected 
five-band apparent magnitudes and their errors, and the redshift $z_i$ if it has a matched spectrum. 
Using the flux and color information, we compute a photometric redshift estimate using a neural 
network technique by \cite{Lin07} as well as $k$-corrections and estimated rest-frame colors for 
each galaxy, which we add as input to the cluster-finding algorithm.

The next step is to define the cluster model we adopt for the filters, including the surface 
density profile $\Sigma_c(r)$, the luminosity function $\phi(M)$, and the assumed Gaussian 
modes of photometric redshift uncertainties. The field density model $P_f(m,z)$ is constructed 
from global number counts of the surveyed background galaxy distributions as a function of 
magnitude and redshift, as shown in equation (4). We then incorporate these models into the Poisson 
likelihood functions as discussed above.

To map the likelihood distributions of the surveyed area, we grid the sky using the Healpix 
package of \cite{Gor05} which provides a useful hierarchical pixelization scheme of equal-area 
pixels. In \citet{Kep99}, the authors choose galaxy positions on an adaptive grid in calculating the 
likelihoods instead of the uniform grids used in the previous matched filter codes \citep{Pos96}, so 
that sufficient resolution in the high density regions is ensured while saving computational 
time and memory for less dense regions. We follow this procedure and evaluate the likelihood 
functions at every galaxy position to locate the peaks in the map as possible cluster centers. 
The cluster richness is optimized over the whole redshift range of our search at intervals that 
finally adapt to $\delta z=0.001$, and for a set of trial cluster scale radii ($r_c$) at $10$ 
kpc $h^{-1}$ steps. The derived quantities for best fit cluster richness, redshift and scale 
length thus correspond to the parameters that maximize the likelihood function at the grid position 
or candidate cluster center.

This algorithm possesses several new features. First, the cluster algorithm is fully adaptive to 
2-D, 2$\half$-D and 3-D case in the optical surveys, and can deal with data with these different 
attributes simultaneously. It can easily accommodate the galaxy redshifts with uncertainties in 
any forms and distributions, from purely single-band imaging data to a complete spectroscopic 
redshift survey, and works well for the intermediate case where photometric redshifts are estimated 
from multi-band color information. Projection effects from foreground--background contamination, 
which have been a long-standing problem for optically-selected clusters, are largely suppressed. 
This allows the detection of even poorer systems at high redshift, and shows great potential for 
current and future large, deeper surveys in the optical band. Second, the current adaptive matched 
filter used a single Poisson statistics in the likelihood analysis, compared to the two-step 
approach in \citet{Kep99}, which uses a ``coarse'' filter based on Gaussian likelihood for 
pre-selection of clusters. We write our code in Fortran-90 and by careful arrangement in computations 
and setting up the quick link search, the optimization of the Poisson likelihood through the 
whole process is now affordable in the sense of execution time and memory. For a survey field of 
$\sim300$ deg$^2$, which is comparable to a typical SDSS stripe \citep{Yor00}, the modified adaptive 
matched filter algorithm requires around 900 megabytes of memory and takes about 30 hours 
for a single run using one dual-processor node in a Linux Beowulf cluster with 3.06 GHz clock speed each. 
With no assumption necessary about sufficiently many galaxies inside each virtual bin as is necessary 
in the Gaussian case, the Poisson 
statistics remains robust in the common situation where there are too few galaxies in each 
cell for Gaussian statistics to apply. Third, as discussed in 
\citet{Whi02} and \citet{Koc03}, the current density model explicitly includes the effect of 
previously found clusters on the global likelihood function. The procedure automatically separates 
overlapping clusters and avoids multiple detections of the same system in the overdensity regions, 
somewhat similar to the CLEAN method used in radio astronomy to produce maps \citep{Hog74, Sch78}. 
We do not need to do extra cluster de-blending work afterwards. Finally, as discussed earlier, our approach to 
maximizing the likelihood differs from most previous cluster-finding techniques that choose a fixed 
cluster scale or search radius. We optimize the core radius for each individual cluster, and the 
cluster richness is computed within a virial radius which is determined from galaxy overdensities. 
This provides insights about the virial mass of such gravitational systems defined by density 
contrast and better corresponds to what is done in theoretical treatments.

\section{Tests on Mock Galaxy Catalogs}\label{sec:Simulation}

To evaluate the completeness and purity (false positive rate) of our
cluster sample, as well as to assess the how well our measured cluster
properties correspond to the properties of the underlying dark matter
halos, we have run the matched-filter algorithm on a mock galaxy catalog
generated from a realistic cosmological N-body simulation. Because of
the large redshift range we are trying to probe, it is important to do
this with as large a simulation volume as possible. In addition,
because we seek here to test the behavior of our algorithm using a
combination of spectroscopic and photometric redshifts, it is useful
to have a realistic galaxy population in both clusters and the field,
with luminosities, colors, and the relation between these quantities
and environment that are a good match to SDSS data.  Here we have used a
mock catalog based on a method namely ADDGALS (Adding Density-Determined 
Galaxies to Lightcone Simulations) 
(\citealt{Wec04} and in preparation, 2007), 
which is designed to model 
relatively bright galaxies in large volume simulations.

The underlying dark matter simulation used here tracks $10^9$ particles
of mass $2.25\times10^{12} h^{-1} M_\odot$ in a periodic cubic volume
with side length of $3 h^{-1}$ Gpc, using a flat $\Lambda$CDM
cosmology with $\Omega_m=0.3$, $\sigma_8=0.9$, and $h=0.7$
\citep[the Hubble Volume simulation;][]{Evr02}. Halos are identified
for masses above $2.7\times10^{13} h^{-1} M_\odot$. Data are collected 
on the past light cone of an observer at the center of the volume.  
The size of the simulation enables the creation of a full-sky survey out 
to redshift of $z=0.58$, and is thus suited to testing our
cluster-finding algorithm out to high redshifts using the SDSS imaging
data.

Galaxies are connected to individual dark matter particles on this
simulated light-cone, subject to several empirical constraints.  The
resolution of the simulation allows the mock catalog to include
galaxies brighter than about 0.4$L^*$; the number of galaxies of a
given brightness placed within the simulation is determined by drawing
galaxies from the SDSS galaxy luminosity function
\citep[][]{Bla03}, with 1.6 mags of luminosity evolution
assumed per unit redshift (the same assumption is made by our cluster
finding algorithm).  The choice of which particle these galaxies are
assigned to is determined by relating the particle overdensities (on a
mass scale of $\sim 1e13 M_{\odot}$) to the two-point correlation
function of the particles; these particles are then chosen to
reproduce the luminosity-dependent correlation function as measured in
the SDSS by \citet[][]{Zeh04}.

Finally, colors are assigned to each galaxy by measuring their local
galaxy density (here, the fifth nearest neighbor within a redshift
slice), and assigning to them the colors of a real SDSS galaxy with
similar luminosity and local density.  The local density measure for
SDSS galaxies is taken from a volume-limited sample of the CMU-Pitt
DR4 Value Added Catalog.  This method produces mock galaxy catalogs
that reproduces the luminosity and color correlation function of the
real sky.  The created mock galaxy sample therefore provides a unique
tool to assess the performance of the SDSS cluster-finding algorithms
in terms of completeness and purity, as well as how the observables of
the detected clusters correspond to dark matter halos assuming galaxy
clusters do trace the underlying halo population in the universe.

Since precise spectroscopic redshift measurements are only available for the SDSS main sample 
galaxies \citep{Str02} and LRGs \citep{Eis01}, we must use photometric redshift estimates for 
most of the galaxies. In order to accurately reproduce this scenario in the simulations, we 
scatter the given redshifts of mock galaxies according to the error distributions of photometric 
redshift estimates, which are obtained by calibrating a sample of $\sim$140,000 SDSS photometric 
redshifts to their known corresponding spectroscopic measurements coming from the SDSS spectroscopic 
survey and various other sources such as CNOC2 \citep{Yee00}, CFRS \citep{Lil95a}, DEEP \citep{Wei05}, 
and 2SLAQ LRG \citep{Pad05}. The photometric redshifts were computed using a neural network 
technique by \cite{Lin07} and in preparation; see also the short discussion in the SDSS DR5 
data release paper, \cite{AMc07}.
The comparison between calculated photometric redshifts and measured spectroscopic redshifts 
is shown in Figure 1 for both the red and blue galaxy samples. The distributions of sampled 
redshift uncertainties are derived for different magnitude and redshift bins, and found to 
be well described by a combination of multiple Gaussian fits as shown in Figure 2 for examples.
The resulted fitting parameters are used for the scattering of mock galaxy redshifts in 
the simulation. In the case of applying the cluster-finding technique to the real SDSS data, however, 
instead of deriving ``empirical'' error estimates collectively, we would use the photo-z errors 
that are computed based on the Nearest Neighbor Error estimate method (NNE) \citep{Lin07}, 
which makes it possible to get an estimate of the error for each individual object. This would 
better constrain the photometric redshift uncertainty, especially for galaxy samples with 
photo-z errors depending strongly on magnitudes and the actual redshifts. We find the 
computed errors correspond reasonably well with the empirical ones derived from statistics, 
with exceptions only for the catastrophic objects. More details would be discussed in a subsequent 
paper on the application of the modified adaptive matched-filter technique with SDSS data.

To summarize, the implementation of simulating the observed galaxy redshifts in the mock sample 
proceeds as follows: for galaxies that satisfy the SDSS spectroscopic target selection criteria 
we take the given galaxy redshifts as spectroscopic measurements, while for the rest of the sample 
we use the scattered redshifts to mimic the photometric redshift estimates. As discussed above 
in \S\ref{sec:Algorithm}, there are a few percent of such galaxies that fall into the redshift 
and magnitude ranges where we find no good calibrations are available. For these galaxies we 
just treat them as if there is no redshift information at all to put into the algorithm. We also 
impose to the mock galaxy catalog an apparent magnitude cut ($r<21$) as we intend to adopt in 
the SDSS imaging sample. The procedure described above thus provides the a mock catalog 
with the most similar characteristics to the SDSS survey and it will allow us to explore the 
performance of the cluster-finding algorithm on real SDSS data.

The modified matched filter algorithm is then run on the mock galaxy catalog, and the 
detected clusters are compared with matched known halos given in the simulation. We find that 
the matches are generally robust against details of the matching techniques, as pointed out 
by \citet[][although see also the discussion of various matching algorithms in \citealt{Roz07}]{Mil05}. 
Here we adopt a matching criterion of projected separation between the detection and the candidate halo 
within the virial radius $r_{200}$ and redshift difference $\Delta z<0.05$. 
To evaluate completeness of the cluster sample, we match each dark halo to the nearest detected cluster 
within the projected cluster $r_{200}$ and $\Delta z$ of 0.05, while in measurement of purity, we 
match clusters to their corresponding halos applying the same criteria. In the case of multiple matches 
which are possible for above matching algorithms, we simply assign the most massive halo within the searching 
space as the real match. Other methods have also been tried in efforts to refine the matching process, 
but no significant changes are found in the final results.

\section{Results and Discussions}\label{sec:Results}

In this section we present the results of running the modified adaptive matched-filter algorithm on 
the simulation-based mock catalogs. These include the completeness and purity check of the detected 
cluster sample, the derived cluster properties such as estimated redshift and richness, and the expected 
scaling relations that would link the observed clusters to true halo distributions.

\subsection{Completeness and Purity Check}\label{sec:Completeness}

We define the completeness $C$ of the selected cluster sample as a cumulative function of $M_{200}$, 
the mass within the virial radius inside which the overdensity is 200 times the critical density:

\begin{equation}
  C(M_{200}) = {N_{found} \over 
    N_{total} }
\end{equation}

where $N_{found}$ is the number of halos with mass greater than $M_{200}$ matched to clusters and 
$N_{total} $ is the total number of halos above that mass.

Figure 3 shows the completeness of the detected cluster sample as a function of redshift and the 
virial mass of matched dark matter halos, respectively. The cluster sample, which has a richness 
cut at $\Lambda_{200} > 20$, is over $95\%$ complete for objects with $M_{200} > 2.0\times10^{14} h^{-1} M_\odot$ 
and $\sim85\%$ complete for objects with masses above $1.0\times10^{14} h^{-1} M_\odot$ in the 
redshift range of $0.05<z<0.45$. As we will find in the subsequent discussion of cluster scaling 
relations, the richness cut we impose on the cluster sample contributes to some of the incompleteness 
for less massive objects because of the large scatter in the cluster 
richness-mass relation; many of the matched clusters at $\sim 1.0\times10^{14} h^{-1} M_\odot$ 
are simply scattered below the richness cut and thus not counted to compute the completeness. This 
can be for sure relieved by lowering the richness cut of the cluster sample, although we choose 
to stick to this cut for the purity considerations below.

Also from Figure 3a, the completeness level of the cluster sample remains almost flat out to 
$z\sim0.45$, beyond which it suffers a significant decline. This is at least partly due to the 
volume limit of the mock catalog which only extends to $z=0.58$. When we scatter 
the given galaxy redshifts with photometric redshift errors, which become large around $z\sim0.5$, 
many of the galaxies near the far edge of the light cone are scattered away while fewer galaxies 
would be shifted into that range, since they are absent from the simulation. The apparent magnitude 
cut we have applied to the mock galaxy sample may also contribute to incompleteness at high redshift. 
Taking into consideration the necessary $k$-corrections, the galaxy sample is no longer complete 
down to the luminosity of $0.4 L^*$, which is the limit assumed throughout the simulation tests. 
The matched filter therefore loses some power in detecting less rich systems at redshifts of $z\sim0.5$ 
and beyond since many fewer galaxies would be bright enough to be observable at that distance in 
the current survey. We have not investigated these effects in detail, though the onset of clear 
incompleteness corresponds well to the distance at which they become important.

We similarly define the purity P of the selected cluster sample as a cumulative function of 
cluster richness $\Lambda_{200}$ which is the total cluster luminosity in units of $L^*$ inside 
its virial radius $r_{200}$

\begin{equation}
  P(\Lambda_{200}) = { N_{match} \over 
    N_{tot,\Lambda}},
\end{equation}

where $ N_{match} $ is the number of clusters with richness greater than $\Lambda_{200}$ matched 
to halos and $N_{tot,\Lambda}$ is the total number of clusters with richness above $\Lambda_{200}$.

The results of the purity check for the obtained cluster catalog are shown in Figure 4. The sample 
is over $95\%$ pure for clusters with $\Lambda_{200} > 30$ and around $90\%$ pure for clusters 
with $\Lambda_{200} > 20$ over the whole redshift range out to $z\sim0.45$. As will be shown 
in the richness-mass relationship below, these two thresholds in richness correspond 
to $M_{200} \sim 6.0\times10^{13} h^{-1} M_\odot$ and $M_{200} \sim 4.0\times10^{13} h^{-1} 
M_\odot$, respectively. It is worth to be noted that the lower purity for $\Lambda_{200} > 20$ 
is clearly going to be affected by halo incompleteness in the simulation, since some of the 
matched halos for this richness will fall below the mass resolution of the halo catalog, which 
means the purity we have derived above is in fact probably a lower limit, in similar logic to 
the completeness arguments.

To ensure a reasonably high purity of selected clusters, we therefore apply a 
$\Lambda_{200} > 20$ cut for the cluster catalog, which is used for analysis of completeness 
as well as cluster derived properties and scaling relations. The purity measurement 
shows a slight but notable uptrend in the last redshift bin of $z\sim0.45-0.5$, which could 
be similarly explained by the arguments above in the completeness discussions. This reflects 
a shift in the richness-mass scaling relation at high redshift end where clusters with the 
same richness measurements may correspond to actually richer and more massive systems because 
of the under-representation of galaxies that are observable in that redshift range. It is 
therefore wise to limit the current cluster catalog to a redshift of $z=0.45$ in order to 
extract a uniform sample for statistical use, though the catalog using real SDSS data may well 
go deeper reliably.

\subsection{Derived Cluster Properties and Scaling Relations}\label{sec:Scaling}

As is discussed in \S\ref{sec:Algorithm}, for each selected cluster a redshift estimate is 
found for the system by the matched filter that optimizes the detection likelihood at the 
given galaxy position as cluster center. This measurement is then taken as the estimated 
redshift for the cluster. Since all the halos have known redshifts in the simulation, by 
matching the detected clusters to halos following the procedure described in 
\S\ref{sec:Simulation} we can compare the derived cluster redshifts with the true 
redshifts of associated halos.

Figure 5 illustrates the comparison between estimated cluster redshifts and known halo 
redshifts. For clusters with redshifts below $z=0.25$ where spectroscopic redshift 
measurements are often available for member galaxies, the derived cluster redshift 
estimates precisely reproduce the true redshifts of corresponding dark halos. 
The inclusion of spectroscopic information of input galaxies markedly sharpens the 
cluster detection likelihood in the line-of-sight dimension and thus provides accurate 
measurements of the cluster redshifts. In the higher redshift range where spectroscopic 
measurements become rare and photometric estimates dominate, the plot illustrates a larger 
dispersion while the matched filter still gives robust determinations of cluster redshifts 
even with only photometric galaxy redshift information for inputs. We find that the 
accuracy of the redshift estimates does incease with cluster richness as expected, which 
is albeit mostly accounted by higher fraction of cluster galaxy members with spectroscopic 
measurements inside these systems. There is a slight uptrend bias seen at the redshift 
of $z\sim0.45$, which we see as a similar indication of incompleteness of the input 
galaxy sample near the high end of the redshift range for this mock catalog because of 
the volume limit and magnitude cut. The estimated cluster redshift determined from maximum 
likelihood tends to drift towards smaller values in some cases since the detection 
probability at higher redshift is suppressed by such effects. We also note the existence 
of a few serious outliers, which probably represent the occasional scenario when there 
exists a mismatch between relevant clusters and dark halos due to the projection effects 
or false positive detections.

The normalized cluster richnesses $\Lambda_{200}$ are also compared with the virial mass 
$M_{200}$ of matched halos. The results are shown in Figure 6. We find that the 
richness-mass scaling relation follows 

\begin{equation}
  \Lambda_{200} = (47.2\pm4.1)\times \left( M_{200}  \over {10^{14} h^{-1} M_\odot} \right )^{1.03\pm0.04} , 
\end{equation}

which is roughly a linear fit. Whether this is correct or not, clearly, depends upon the details
of the simulation input, and the way the simulation was constructed gives no easy clue to
what the results should be. What is important in this test, however, is that we recover
what is present in the simulations, not what might or might not be present in the real universe.
To that end, we have constructed three more plots. The first, Figure 7, compares the cluster
richness determined by the present algorithm with the total three dimensional luminosity of the
matched halos; the agreement is very good, with no bias evident at either the sparse or the
rich end. Given this agreement and the results of Figure 6, the next plot, Figure 8, of the 3-D halo 
luminosity vs the 3-D halo mass, contains no surprises. The simulated halo mass is, in fact,
linear with its total luminosity, and we recover this relationship.

Figure 9 compares the derived cluster virial radius $r_{200}$ from the cluster-finding algorithm 
and the $r_{200}$ determined from 3-dimensional galaxy overdensities. The agreement is excellent 
at small virial radii, though there is a strong hint that the
algorithm slightly overestimates large virial radii, by seven percent or thereabouts. This is
almost certainly due to the assumption of a single NFW profile to describe the cluster; neighboring
halos have rather different effects in the cylinder to which the algorithm is sensitive and the
corresponding sphere in the simulations, but it is gratifying that the effects are this small.
These results further justify our choice to refer our richness measurements 
to the commonly-used virial radius determined from galaxy overdensities.

It is, however, clear that the scatter in the richness--mass relation derived from the cluster 
finding algorithm (Figure 6) is somewhat larger than that of the intrinsic richeness-mass relation 
in the simulations (Figure 8), 
which can be read as an indication of complications in the 
cluster-halo matching process, e.g., the inevitable difference between the cluster finder 
and halo finder regarding fragmentation and merging, differing shapes between the galaxy 
and mass distributions, and, even further, the variable mass-to-light ratios inside the systems 
incorporated in the current dark matter simulations. Despite these intrinsic 
dispersions, the richness-mass scaling relation shows a strong linear correspondence 
between the observables and the mass, and thus makes it possible to extract 
the true halo distribution in the Universe from the observed cluster abundance 
and correlation functions. It is important to note that the simulation from which
the catalog was made is a dark-matter-only simulation, and thus effects which
may well exist in real clusters and can affect the baryon fraction in the
intracluster gas and galaxies (see, for example, \cite{Kra05}) as a function of
cluster mass are absent here, but the fact that we recover the relation found from 
input 3-D simulations, here just linear, indicates that we should be able to investigate
a possibly more complex relationship in the real universe.

\section{Conclusions}\label{sec:Conclusion}

We present a modified matched filter algorithm which is designed to construct a 
comprehensive cluster catalog from the Sloan Digital Sky Survey, but is applicable
to any deep photometric survey. The technique is fully adaptive to 2-D, 2$\half$-D 
and 3-D optical surveys, as well as to various cluster scales and substructures.

The cluster-finding algorithm has been tested against a realistic mock SDSS catalog from 
a large N-body simulation. The results suggest that the selected cluster sample is 
$\sim 85\%$ complete and over $90\%$ pure for systems more massive than 
$1.0\times10^{14} h^{-1}$ M$_\odot$ with redshifts up to $z=0.45$. The estimated 
cluster redshifts derived from maximum likelihood analysis show small errors 
with $\Delta z < 0.01$, and the normalized cluster richness measurements fit 
linearly with the virial mass of matched halos, the correct relation in this simulation.
This offers hope that the (very likely nonlinear) relation between richness
and halo mass which exists in the real universe can be investigated with these
techniques.

\acknowledgments F.D. thanks H. Lin, H. Oyaizu, and the SDSS photo-$z$
group for providing the photometric redshifts which allowed us to
derive the statistics of the photo-$z$ calibration to the
spectroscopic redshifts. E.P. is an ADVANCE fellow (NSF grant
AST-0649899), also supported by NASA grant NAG5-11489.  RHW was
supported in part by the U.S. Department of Energy under contract
number DE-AC02-76SF00515.  This research used 
computational facilities supported by NSF grant AST-0216105.

\bibliography{ms.bib}

\begin{thebibliography}{92}
\expandafter\ifx\csname natexlab\endcsname\relax\def\natexlab#1{#1}\fi

\bibitem[{{Abell}(1958)}]{Abe58}
{Abell}, G.~O. 1958, \apjs, 3, 211

\bibitem[{{Abell} {et~al.}(1989){Abell}, {Corwin}, \& {Olowin}}]{Abe89}
{Abell}, G.~O., {Corwin}, H.~G., \& {Olowin}, R.~P. 1989, \apjs, 70, 1

\bibitem[{{Adelman-McCarthy}(2007)}]{AMc07}
{Adelman-McCarthy}, J.~K, e.~a. 2007, \apjs, accepted for publication

\bibitem[{{Annis} {et~al.}(2002){Annis}, {Makler}, {Kent}, {Dodelson},
  {Frieman}, {Sheldon}, {McKay}, {Bahcall}, \& {SDSS Collaboration}}]{Ann02}
{Annis}, J., {Makler}, M., {Kent}, S., {Dodelson}, S., {Frieman}, J.,
  {Sheldon}, E., {McKay}, T., {Bahcall}, N., \& {SDSS Collaboration}. 2002, in
  Bulletin of the American Astronomical Society, 778--+

\bibitem[{{B{\" o}hringer} {et~al.}(2001){B{\" o}hringer}, {Schuecker},
  {Guzzo}, {Collins}, {Voges}, {Schindler}, {Neumann}, {Cruddace}, {De Grandi},
  {Chincarini}, {Edge}, {MacGillivray}, \& {Shaver}}]{Boh01}
{B{\" o}hringer}, H., {Schuecker}, P., {Guzzo}, L., {Collins}, C.~A., {Voges},
  W., {Schindler}, S., {Neumann}, D.~M., {Cruddace}, R.~G., {De Grandi}, S.,
  {Chincarini}, G., {Edge}, A.~C., {MacGillivray}, H.~T., \& {Shaver}, P. 2001,
  \aap, 369, 826

\bibitem[{{Bahcall}(1988)}]{Bah88}
{Bahcall}, N.~A. 1988, \araa, 26, 631

\bibitem[{{Bahcall} {et~al.}(2003){Bahcall}, {McKay}, {Annis}, {Kim}, {Dong},
  {Hansen}, {Goto}, {Gunn}, {Miller}, {Nichol}, {Postman}, {Schneider},
  {Schroeder}, {Voges}, {Brinkmann}, \& {Fukugita}}]{Bah03}
{Bahcall}, N.~A., {McKay}, T.~A., {Annis}, J., {Kim}, R.~S.~J., {Dong}, F.,
  {Hansen}, S., {Goto}, T., {Gunn}, J.~E., {Miller}, C., {Nichol}, R.~C.,
  {Postman}, M., {Schneider}, D., {Schroeder}, J., {Voges}, W., {Brinkmann},
  J., \& {Fukugita}, M. 2003, \apjs, 148, 243

\bibitem[{{Bahcall} {et~al.}(1999){Bahcall}, {Ostriker}, {Perlmutter}, \&
  {Steinhardt}}]{Bah99}
{Bahcall}, N.~A., {Ostriker}, J.~P., {Perlmutter}, S., \& {Steinhardt}, P.~J.
  1999, Science, 284, 1481

\bibitem[{{Baldry} {et~al.}(2005){Baldry}, {Glazebrook}, {Budav{\'a}ri},
  {Eisenstein}, {Annis}, {Bahcall}, {Blanton}, {Brinkmann}, {Csabai},
  {Heckman}, {Lin}, {Loveday}, {Nichol}, \& {Schneider}}]{Bal05}
{Baldry}, I.~K., {Glazebrook}, K., {Budav{\'a}ri}, T., {Eisenstein}, D.~J.,
  {Annis}, J., {Bahcall}, N.~A., {Blanton}, M.~R., {Brinkmann}, J., {Csabai},
  I., {Heckman}, T.~M., {Lin}, H., {Loveday}, J., {Nichol}, R.~C., \&
  {Schneider}, D.~P. 2005, \mnras, 358, 441

\bibitem[{{Bartelmann}(1996)}]{Bar96}
{Bartelmann}, M. 1996, \aap, 313, 697

\bibitem[{{Blanton} {et~al.}(2003){Blanton}, {Hogg}, {Bahcall}, {Brinkmann},
  {Britton}, {Connolly}, {Csabai}, {Fukugita}, {Loveday}, {Meiksin}, {Munn},
  {Nichol}, {Okamura}, {Quinn}, {Schneider}, {Shimasaku}, {Strauss}, {Tegmark},
  {Vogeley}, \& {Weinberg}}]{Bla03}
{Blanton}, M.~R., {Hogg}, D.~W., {Bahcall}, N.~A., {Brinkmann}, J., {Britton},
  M., {Connolly}, A.~J., {Csabai}, I., {Fukugita}, M., {Loveday}, J.,
  {Meiksin}, A., {Munn}, J.~A., {Nichol}, R.~C., {Okamura}, S., {Quinn}, T.,
  {Schneider}, D.~P., {Shimasaku}, K., {Strauss}, M.~A., {Tegmark}, M.,
  {Vogeley}, M.~S., \& {Weinberg}, D.~H. 2003, \apj, 592, 819

\bibitem[{{B{\"o}hringer} {et~al.}(2004){B{\"o}hringer}, {Schuecker}, {Guzzo},
  {Collins}, {Voges}, {Cruddace}, {Ortiz-Gil}, {Chincarini}, {De Grandi},
  {Edge}, {MacGillivray}, {Neumann}, {Schindler}, \& {Shaver}}]{Boh04}
{B{\"o}hringer}, H., {Schuecker}, P., {Guzzo}, L., {Collins}, C.~A., {Voges},
  W., {Cruddace}, R.~G., {Ortiz-Gil}, A., {Chincarini}, G., {De Grandi}, S.,
  {Edge}, A.~C., {MacGillivray}, H.~T., {Neumann}, D.~M., {Schindler}, S., \&
  {Shaver}, P. 2004, \aap, 425, 367

\bibitem[{{Carlstrom} {et~al.}(2000){Carlstrom}, {Joy}, {Grego}, {Holder},
  {Holzapfel}, {Mohr}, {Patel}, \& {Reese}}]{Car00}
{Carlstrom}, J.~E., {Joy}, M.~K., {Grego}, L., {Holder}, G.~P., {Holzapfel},
  W.~L., {Mohr}, J.~J., {Patel}, S., \& {Reese}, E.~D. 2000, Physica Scripta
  Volume T, 85, 148

\bibitem[{{Christlein}(2000)}]{Chr00}
{Christlein}, D. 2000, \apj, 545, 145

\bibitem[{{Croton} {et~al.}(2005){Croton}, {Farrar}, {Norberg}, {Colless},
  {Peacock}, {Baldry}, {Baugh}, {Bland-Hawthorn}, {Bridges}, {Cannon}, {Cole},
  {Collins}, {Couch}, {Dalton}, {De Propris}, {Driver}, {Efstathiou}, {Ellis},
  {Frenk}, {Glazebrook}, {Jackson}, {Lahav}, {Lewis}, {Lumsden}, {Maddox},
  {Madgwick}, {Peterson}, {Sutherland}, \& {Taylor}}]{Cro05}
{Croton}, D.~J., {Farrar}, G.~R., {Norberg}, P., {Colless}, M., {Peacock},
  J.~A., {Baldry}, I.~K., {Baugh}, C.~M., {Bland-Hawthorn}, J., {Bridges}, T.,
  {Cannon}, R., {Cole}, S., {Collins}, C., {Couch}, W., {Dalton}, G., {De
  Propris}, R., {Driver}, S.~P., {Efstathiou}, G., {Ellis}, R.~S., {Frenk},
  C.~S., {Glazebrook}, K., {Jackson}, C., {Lahav}, O., {Lewis}, I., {Lumsden},
  S., {Maddox}, S., {Madgwick}, D., {Peterson}, B.~A., {Sutherland}, W., \&
  {Taylor}, K. 2005, \mnras, 356, 1155

\bibitem[{{Dalton} {et~al.}(1994){Dalton}, {Efstathiou}, {Maddox}, \&
  {Sutherland}}]{Dal94}
{Dalton}, G.~B., {Efstathiou}, G., {Maddox}, S.~J., \& {Sutherland}, W.~J.
  1994, \mnras, 269, 151

\bibitem[{{Dalton} {et~al.}(1997){Dalton}, {Maddox}, {Sutherland}, \&
  {Efstathiou}}]{Dal97}
{Dalton}, G.~B., {Maddox}, S.~J., {Sutherland}, W.~J., \& {Efstathiou}, G.
  1997, \mnras, 289, 263

\bibitem[{{Dressler}(1980)}]{Dre80}
{Dressler}, A. 1980, \apj, 236, 351

\bibitem[{{Dressler}(1984)}]{Dre84}
---. 1984, \araa, 22, 185

\bibitem[{{Dressler} \& {Gunn}(1992)}]{Dre92}
{Dressler}, A. \& {Gunn}, J.~E. 1992, \apjs, 78, 1

\bibitem[{{Ebeling} {et~al.}(1998){Ebeling}, {Edge}, {Bohringer}, {Allen},
  {Crawford}, {Fabian}, {Voges}, \& {Huchra}}]{Ebe98}
{Ebeling}, H., {Edge}, A.~C., {Bohringer}, H., {Allen}, S.~W., {Crawford},
  C.~S., {Fabian}, A.~C., {Voges}, W., \& {Huchra}, J.~P. 1998, \mnras, 301,
  881

\bibitem[{{Edge} {et~al.}(1990){Edge}, {Stewart}, {Fabian}, \&
  {Arnaud}}]{Edg90}
{Edge}, A.~C., {Stewart}, G.~C., {Fabian}, A.~C., \& {Arnaud}, K.~A. 1990,
  \mnras, 245, 559

\bibitem[{{Eisenstein} {et~al.}(2001){Eisenstein}, {Annis}, {Gunn}, {Szalay},
  {Connolly}, {Nichol}, {Bahcall}, {Bernardi}, {Burles}, {Castander},
  {Fukugita}, {Hogg}, {Ivezi{\' c}}, {Knapp}, {Lupton}, {Narayanan}, {Postman},
  {Reichart}, {Richmond}, {Schneider}, {Schlegel}, {Strauss}, {SubbaRao},
  {Tucker}, {Vanden Berk}, {Vogeley}, {Weinberg}, \& {Yanny}}]{Eis01}
{Eisenstein}, D.~J., {Annis}, J., {Gunn}, J.~E., {Szalay}, A.~S., {Connolly},
  A.~J., {Nichol}, R.~C., {Bahcall}, N.~A., {Bernardi}, M., {Burles}, S.,
  {Castander}, F.~J., {Fukugita}, M., {Hogg}, D.~W., {Ivezi{\' c}}, {\v Z}.,
  {Knapp}, G.~R., {Lupton}, R.~H., {Narayanan}, V., {Postman}, M., {Reichart},
  D.~E., {Richmond}, M., {Schneider}, D.~P., {Schlegel}, D.~J., {Strauss},
  M.~A., {SubbaRao}, M., {Tucker}, D.~L., {Vanden Berk}, D., {Vogeley}, M.~S.,
  {Weinberg}, D.~H., \& {Yanny}, B. 2001, \aj, 122, 2267

\bibitem[{{Evrard}(1989)}]{Evr89}
{Evrard}, A.~E. 1989, \apjl, 341, L71

\bibitem[{{Evrard} {et~al.}(2002){Evrard}, {MacFarland}, {Couchman}, {Colberg},
  {Yoshida}, {White}, {Jenkins}, {Frenk}, {Pearce}, {Peacock}, \&
  {Thomas}}]{Evr02}
{Evrard}, A.~E., {MacFarland}, T.~J., {Couchman}, H.~M.~P., {Colberg}, J.~M.,
  {Yoshida}, N., {White}, S.~D.~M., {Jenkins}, A., {Frenk}, C.~S., {Pearce},
  F.~R., {Peacock}, J.~A., \& {Thomas}, P.~A. 2002, \apj, 573, 7

\bibitem[{{Folkes} {et~al.}(1999){Folkes}, {Ronen}, {Price}, {Lahav},
  {Colless}, {Maddox}, {Deeley}, {Glazebrook}, {Bland-Hawthorn}, {Cannon},
  {Cole}, {Collins}, {Couch}, {Driver}, {Dalton}, {Efstathiou}, {Ellis},
  {Frenk}, {Kaiser}, {Lewis}, {Lumsden}, {Peacock}, {Peterson}, {Sutherland},
  \& {Taylor}}]{Fol99}
{Folkes}, S., {Ronen}, S., {Price}, I., {Lahav}, O., {Colless}, M., {Maddox},
  S., {Deeley}, K., {Glazebrook}, K., {Bland-Hawthorn}, J., {Cannon}, R.,
  {Cole}, S., {Collins}, C., {Couch}, W., {Driver}, S.~P., {Dalton}, G.,
  {Efstathiou}, G., {Ellis}, R.~S., {Frenk}, C.~S., {Kaiser}, N., {Lewis}, I.,
  {Lumsden}, S., {Peacock}, J., {Peterson}, B.~A., {Sutherland}, W., \&
  {Taylor}, K. 1999, \mnras, 308, 459

\bibitem[{{Fukugita} {et~al.}(1996){Fukugita}, {Ichikawa}, {Gunn}, {Doi},
  {Shimasaku}, \& {Schneider}}]{Fuk96}
{Fukugita}, M., {Ichikawa}, T., {Gunn}, J.~E., {Doi}, M., {Shimasaku}, K., \&
  {Schneider}, D.~P. 1996, \aj, 111, 1748

\bibitem[{{Geller} \& {Huchra}(1983)}]{Gel83}
{Geller}, M.~J. \& {Huchra}, J.~P. 1983, \apjs, 52, 61

\bibitem[{{Gioia} {et~al.}(1990){Gioia}, {Maccacaro}, {Schild}, {Wolter},
  {Stocke}, {Morris}, \& {Henry}}]{Gio90}
{Gioia}, I.~M., {Maccacaro}, T., {Schild}, R.~E., {Wolter}, A., {Stocke},
  J.~T., {Morris}, S.~L., \& {Henry}, J.~P. 1990, \apjs, 72, 567

\bibitem[{{Gladders} \& {Yee}(2000)}]{Gla00}
{Gladders}, M.~D. \& {Yee}, H.~K.~C. 2000, \aj, 120, 2148

\bibitem[{{Gladders} \& {Yee}(2005)}]{Gla05}
---. 2005, \apjs, 157, 1

\bibitem[{{G{\'o}rski} {et~al.}(2005){G{\'o}rski}, {Hivon}, {Banday},
  {Wandelt}, {Hansen}, {Reinecke}, \& {Bartelmann}}]{Gor05}
{G{\'o}rski}, K.~M., {Hivon}, E., {Banday}, A.~J., {Wandelt}, B.~D., {Hansen},
  F.~K., {Reinecke}, M., \& {Bartelmann}, M. 2005, \apj, 622, 759

\bibitem[{{Goto} {et~al.}(2002){Goto}, {Sekiguchi}, {Nichol}, {Bahcall}, {Kim},
  {Annis}, {Ivezi{\' c}}, {Brinkmann}, {Hennessy}, {Szokoly}, \&
  {Tucker}}]{Got02}
{Goto}, T., {Sekiguchi}, M., {Nichol}, R.~C., {Bahcall}, N.~A., {Kim},
  R.~S.~J., {Annis}, J., {Ivezi{\' c}}, {\v Z}., {Brinkmann}, J., {Hennessy},
  G.~S., {Szokoly}, G.~P., \& {Tucker}, D.~L. 2002, \aj, 123, 1807

\bibitem[{{Gunn} {et~al.}(1998){Gunn}, {Carr}, {Rockosi}, {Sekiguchi}, {Berry},
  {Elms}, {de Haas}, {Ivezi{\'c}}, {Knapp}, {Lupton}, {Pauls}, {Simcoe},
  {Hirsch}, {Sanford}, {Wang}, {York}, {Harris}, {Annis}, {Bartozek},
  {Boroski}, {Bakken}, {Haldeman}, {Kent}, {Holm}, {Holmgren}, {Petravick},
  {Prosapio}, {Rechenmacher}, {Doi}, {Fukugita}, {Shimasaku}, {Okada}, {Hull},
  {Siegmund}, {Mannery}, {Blouke}, {Heidtman}, {Schneider}, {Lucinio}, \&
  {Brinkman}}]{Gun98}
{Gunn}, J.~E., {Carr}, M., {Rockosi}, C., {Sekiguchi}, M., {Berry}, K., {Elms},
  B., {de Haas}, E., {Ivezi{\'c}}, {\v Z}., {Knapp}, G., {Lupton}, R., {Pauls},
  G., {Simcoe}, R., {Hirsch}, R., {Sanford}, D., {Wang}, S., {York}, D.,
  {Harris}, F., {Annis}, J., {Bartozek}, L., {Boroski}, W., {Bakken}, J.,
  {Haldeman}, M., {Kent}, S., {Holm}, S., {Holmgren}, D., {Petravick}, D.,
  {Prosapio}, A., {Rechenmacher}, R., {Doi}, M., {Fukugita}, M., {Shimasaku},
  K., {Okada}, N., {Hull}, C., {Siegmund}, W., {Mannery}, E., {Blouke}, M.,
  {Heidtman}, D., {Schneider}, D., {Lucinio}, R., \& {Brinkman}, J. 1998, \aj,
  116, 3040

\bibitem[{{Hansen} {et~al.}(2005){Hansen}, {McKay}, {Wechsler}, {Annis},
  {Sheldon}, \& {Kimball}}]{Han05}
{Hansen}, S.~M., {McKay}, T.~A., {Wechsler}, R.~H., {Annis}, J., {Sheldon},
  E.~S., \& {Kimball}, A. 2005, \apj, 633, 122

\bibitem[{{Henry}(2000)}]{Hen00}
{Henry}, J.~P. 2000, \apj, 534, 565

\bibitem[{{H{\"o}gbom}(1974)}]{Hog74} 
{H{\"o}gbom}, J.~A.\ 1974, \aaps, 15, 417 

\bibitem[{{Hogg} {et~al.}(2003){Hogg}, {Blanton}, {Eisenstein}, {Gunn},
  {Schlegel}, {Zehavi}, {Bahcall}, {Brinkmann}, {Csabai}, {Schneider},
  {Weinberg}, \& {York}}]{Hog03}
{Hogg}, D.~W., {Blanton}, M.~R., {Eisenstein}, D.~J., {Gunn}, J.~E.,
  {Schlegel}, D.~J., {Zehavi}, I., {Bahcall}, N.~A., {Brinkmann}, J., {Csabai},
  I., {Schneider}, D.~P., {Weinberg}, D.~H., \& {York}, D.~G. 2003, \apjl, 585,
  L5

\bibitem[{{Huchra} \& {Geller}(1982)}]{Huc82}
{Huchra}, J.~P. \& {Geller}, M.~J. 1982, \apj, 257, 423

\bibitem[{{Ilbert} {et~al.}(2005){Ilbert}, {Tresse}, {Zucca}, {Bardelli},
  {Arnouts}, {Zamorani}, {Pozzetti}, {Bottini}, {Garilli}, {Le Brun}, {Le
  F{\`e}vre}, {Maccagni}, {Picat}, {Scaramella}, {Scodeggio}, {Vettolani},
  {Zanichelli}, {Adami}, {Arnaboldi}, {Bolzonella}, {Cappi}, {Charlot},
  {Contini}, {Foucaud}, {Franzetti}, {Gavignaud}, {Guzzo}, {Iovino},
  {McCracken}, {Marano}, {Marinoni}, {Mathez}, {Mazure}, {Meneux}, {Merighi},
  {Paltani}, {Pello}, {Pollo}, {Radovich}, {Bondi}, {Bongiorno}, {Busarello},
  {Ciliegi}, {Lamareille}, {Mellier}, {Merluzzi}, {Ripepi}, \& {Rizzo}}]{Ilb05}
{Ilbert}, O., {Tresse}, L., {Zucca}, E., {Bardelli}, S., {Arnouts}, S.,
  {Zamorani}, G., {Pozzetti}, L., {Bottini}, D., {Garilli}, B., {Le Brun}, V.,
  {Le F{\`e}vre}, O., {Maccagni}, D., {Picat}, J.-P., {Scaramella}, R.,
  {Scodeggio}, M., {Vettolani}, G., {Zanichelli}, A., {Adami}, C., {Arnaboldi},
  M., {Bolzonella}, M., {Cappi}, A., {Charlot}, S., {Contini}, T., {Foucaud},
  S., {Franzetti}, P., {Gavignaud}, I., {Guzzo}, L., {Iovino}, A., {McCracken},
  H.~J., {Marano}, B., {Marinoni}, C., {Mathez}, G., {Mazure}, A., {Meneux},
  B., {Merighi}, R., {Paltani}, S., {Pello}, R., {Pollo}, A., {Radovich}, M.,
  {Bondi}, M., {Bongiorno}, A., {Busarello}, G., {Ciliegi}, P., {Lamareille},
  F., {Mellier}, Y., {Merluzzi}, P., {Ripepi}, V., \& {Rizzo}, D. 2005, \aap,
  439, 863

\bibitem[{{Kawasaki} {et~al.}(1998){Kawasaki}, {Shimasaku}, {Doi}, \&
  {Okamura}}]{Kaw98}
{Kawasaki}, W., {Shimasaku}, K., {Doi}, M., \& {Okamura}, S. 1998, \aaps, 130,
  567

\bibitem[{{Kepner} {et~al.}(1999){Kepner}, {Fan}, {Bahcall}, {Gunn}, {Lupton},
  \& {Xu}}]{Kep99}
{Kepner}, J., {Fan}, X., {Bahcall}, N., {Gunn}, J., {Lupton}, R., \& {Xu}, G.
  1999, \apj, 517, 78

\bibitem[{{Kim} {et~al.}(2002){Kim}, {Kepner}, {Postman}, {Strauss}, {Bahcall},
  {Gunn}, {Lupton}, {Annis}, {Nichol}, {Castander}, {Brinkmann}, {Brunner},
  {Connolly}, {Csabai}, {Hindsley}, {Ivezi{\' c}}, {Vogeley}, \&
  {York}}]{Kim02}
{Kim}, R.~S.~J., {Kepner}, J.~V., {Postman}, M., {Strauss}, M.~A., {Bahcall},
  N.~A., {Gunn}, J.~E., {Lupton}, R.~H., {Annis}, J., {Nichol}, R.~C.,
  {Castander}, F.~J., {Brinkmann}, J., {Brunner}, R.~J., {Connolly}, A.,
  {Csabai}, I., {Hindsley}, R.~B., {Ivezi{\' c}}, {\v Z}., {Vogeley}, M.~S., \&
  {York}, D.~G. 2002, \aj, 123, 20

\bibitem[{{Kochanek} {et~al.}(2003){Kochanek}, {White}, {Huchra}, {Macri},
  {Jarrett}, {Schneider}, \& {Mader}}]{Koc03}
{Kochanek}, C.~S., {White}, M., {Huchra}, J., {Macri}, L., {Jarrett}, T.~H.,
  {Schneider}, S.~E., \& {Mader}, J. 2003, \apj, 585, 161

\bibitem[{{Koester} {et~al.}(2007){Koester}, {McKay}, {Annis}, {Wechsler},
  {Evrard}, {Bleem}, {Becker}, {Johnston}, {Sheldon}, {Nichol}, {Miller},
  {Scranton}, {Bahcall}, {Barentine}, {Brewington}, {Brinkmann}, {Harvanek},
  {Kleinman}, {Krzesinski}, {Long}, {Nitta}, {Schneider}, {Sneddin}, {Voges},
  {York}, \& {SDSS collaboration}}]{Koe07}
{Koester}, B.~P., {McKay}, T.~A., {Annis}, J., {Wechsler}, R.~H., {Evrard}, A.,
  {Bleem}, L., {Becker}, M., {Johnston}, D., {Sheldon}, E., {Nichol}, R.,
  {Miller}, C., {Scranton}, R., {Bahcall}, N., {Barentine}, J., {Brewington},
  H., {Brinkmann}, J., {Harvanek}, M., {Kleinman}, S., {Krzesinski}, J.,
  {Long}, D., {Nitta}, A., {Schneider}, D., {Sneddin}, S., {Voges}, W., {York},
  D., \& {SDSS collaboration}. 2007, ArXiv Astrophysics e-prints

\bibitem[{{Kravtsov} {et~al.}(2005){Kravtsov}, {Nagai}, \& {Vikhlinin}}]{Kra05}
{Kravtsov}, A.~V., {Nagai}, D., \& {Vikhlinin}, A.~A. 2005, \apj, 625, 588

\bibitem[{{Lilly} {et~al.}(1995{\natexlab{a}}){Lilly}, {Le Fevre}, {Crampton},
  {Hammer}, \& {Tresse}}]{Lil95a}
{Lilly}, S.~J., {Le Fevre}, O., {Crampton}, D., {Hammer}, F., \& {Tresse}, L.
  1995{\natexlab{a}}, \apj, 455, 50

\bibitem[{{Lilly} {et~al.}(1995{\natexlab{b}}){Lilly}, {Tresse}, {Hammer},
  {Crampton}, \& {Le Fevre}}]{Lil95b}
{Lilly}, S.~J., {Tresse}, L., {Hammer}, F., {Crampton}, D., \& {Le Fevre}, O.
  1995{\natexlab{b}}, \apj, 455, 108

\bibitem[{{Lin} {et~al.}(2006){Lin}, {Lima}, {Oyaizu}, {Cunha}, {Frieman},
  {Annis}, {Koester}, {Hao}, {McKay}, \& {Sheldon}}]{Lin07}
{Lin}, H., {Lima}, M., {Oyaizu}, H., {Cunha}, C., {Frieman}, J., {Annis}, J.,
  {Koester}, B., {Hao}, J., {McKay}, T., \& {Sheldon}, E. 2006, in American
  Astronomical Society Meeting Abstracts, 215.03--+

\bibitem[{{Lin} {et~al.}(1999){Lin}, {Yee}, {Carlberg}, {Morris}, {Sawicki},
  {Patton}, {Wirth}, \& {Shepherd}}]{Lin99}
{Lin}, H., {Yee}, H.~K.~C., {Carlberg}, R.~G., {Morris}, S.~L., {Sawicki}, M.,
  {Patton}, D.~R., {Wirth}, G., \& {Shepherd}, C.~W. 1999, \apj, 518, 533

\bibitem[{{Lin} \& {Mohr}(2004)}]{Lin04b}
{Lin}, Y.-T. \& {Mohr}, J.~J. 2004, \apj, 617, 879

\bibitem[{{Lin} \& {Mohr}(2007)}]{Lin07a} 
{Lin}, Y.-T., \& {Mohr}, J.~J.\ 2007, \apjs, 170, 71 

\bibitem[{{Lin} {et~al.}(2004){Lin}, {Mohr}, \& {Stanford}}]{Lin04a}
{Lin}, Y.-T., {Mohr}, J.~J., \& {Stanford}, S.~A. 2004, \apj, 610, 745

\bibitem[{{Loveday}(2004)}]{Lov04}
{Loveday}, J. 2004, \mnras, 347, 601

\bibitem[{{Loveday} {et~al.}(1992){Loveday}, {Peterson}, {Efstathiou}, \&
  {Maddox}}]{Lov92}
{Loveday}, J., {Peterson}, B.~A., {Efstathiou}, G., \& {Maddox}, S.~J. 1992,
  \apj, 390, 338

\bibitem[{{Lumsden} {et~al.}(1992){Lumsden}, {Nichol}, {Collins}, \&
  {Guzzo}}]{Lum92}
{Lumsden}, S.~L., {Nichol}, R.~C., {Collins}, C.~A., \& {Guzzo}, L. 1992,
  \mnras, 258, 1

\bibitem[{{Lupton} {et~al.}(2001){Lupton}, {Gunn}, {Ivezi{\'c}}, {Knapp},
  {Kent}, \& {Yasuda}}]{Lup01}
{Lupton}, R., {Gunn}, J.~E., {Ivezi{\'c}}, Z., {Knapp}, G.~R., {Kent}, S., \&
  {Yasuda}, N. 2001, in ASP Conf. Ser. 238: Astronomical Data Analysis Software
  and Systems X, ed. F.~R. {Harnden}, Jr., F.~A. {Primini}, \& H.~E. {Payne},
  269--+

\bibitem[{{Miller} {et~al.}(2005){Miller}, {Nichol}, {Reichart}, {Wechsler},
  {Evrard}, {Annis}, {McKay}, {Bahcall}, {Bernardi}, {Boehringer}, {Connolly},
  {Goto}, {Kniazev}, {Lamb}, {Postman}, {Schneider}, {Sheth}, \&
  {Voges}}]{Mil05}
{Miller}, C.~J., {Nichol}, R.~C., {Reichart}, D., {Wechsler}, R.~H., {Evrard},
  A.~E., {Annis}, J., {McKay}, T.~A., {Bahcall}, N.~A., {Bernardi}, M.,
  {Boehringer}, H., {Connolly}, A.~J., {Goto}, T., {Kniazev}, A., {Lamb}, D.,
  {Postman}, M., {Schneider}, D.~P., {Sheth}, R.~K., \& {Voges}, W. 2005, \aj,
  130, 968

\bibitem[{{Mo} {et~al.}(2004){Mo}, {Yang}, {van den Bosch}, \& {Jing}}]{Mo04}
{Mo}, H.~J., {Yang}, X., {van den Bosch}, F.~C., \& {Jing}, Y.~P. 2004, \mnras,
  349, 205

\bibitem[{{Mohr} {et~al.}(2002){Mohr}, {Carlstrom}, \& {The Sza
  Collaboration}}]{Moh02}
{Mohr}, J.~J., {Carlstrom}, J.~E., \& {The Sza Collaboration}. 2002, in ASP
  Conf. Ser. 257: AMiBA 2001: High-Z Clusters, Missing Baryons, and CMB
  Polarization, 43--+

\bibitem[{{Mullis} {et~al.}(2003){Mullis}, {McNamara}, {Quintana}, {Vikhlinin},
  {Henry}, {Gioia}, {Hornstrup}, {Forman}, \& {Jones}}]{Mul03}
{Mullis}, C.~R., {McNamara}, B.~R., {Quintana}, H., {Vikhlinin}, A., {Henry},
  J.~P., {Gioia}, I.~M., {Hornstrup}, A., {Forman}, W., \& {Jones}, C. 2003,
  \apj, 594, 154

\bibitem[{{Nagai} \& {Kravtsov}(2005)}]{Nag05}
{Nagai}, D. \& {Kravtsov}, A.~V. 2005, \apj, 618, 557

\bibitem[{{Nagamine} {et~al.}(2001){Nagamine}, {Fukugita}, {Cen}, \&
  {Ostriker}}]{Nag01}
{Nagamine}, K., {Fukugita}, M., {Cen}, R., \& {Ostriker}, J.~P. 2001, \mnras,
  327, L10

\bibitem[{{Navarro} {et~al.}(1996){Navarro}, {Frenk}, \& {White}}]{Nav96}
{Navarro}, J.~F., {Frenk}, C.~S., \& {White}, S.~D.~M. 1996, \apj, 462, 563

\bibitem[{{Padmanabhan} {et~al.}(2005){Padmanabhan}, {Budav{\'a}ri},
  {Schlegel}, {Bridges}, {Brinkmann}, {Cannon}, {Connolly}, {Croom}, {Csabai},
  {Drinkwater}, {Eisenstein}, {Hewett}, {Loveday}, {Nichol}, {Pimbblet}, {De
  Propris}, {Schneider}, {Scranton}, {Seljak}, {Shanks}, {Szapudi}, {Szalay},
  \& {Wake}}]{Pad05}
{Padmanabhan}, N., {Budav{\'a}ri}, T., {Schlegel}, D.~J., {Bridges}, T.,
  {Brinkmann}, J., {Cannon}, R., {Connolly}, A.~J., {Croom}, S.~M., {Csabai},
  I., {Drinkwater}, M., {Eisenstein}, D.~J., {Hewett}, P.~C., {Loveday}, J.,
  {Nichol}, R.~C., {Pimbblet}, K.~A., {De Propris}, R., {Schneider}, D.~P.,
  {Scranton}, R., {Seljak}, U., {Shanks}, T., {Szapudi}, I., {Szalay}, A.~S.,
  \& {Wake}, D. 2005, \mnras, 359, 237

\bibitem[{{Pierpaoli} {et~al.}(2005){Pierpaoli}, {Anthoine}, {Huffenberger}, \&
  {Daubechies}}]{Pierpa05}
{Pierpaoli}, E., {Anthoine}, S., {Huffenberger}, K., \& {Daubechies}, I. 2005,
  \mnras, 359, 261

\bibitem[{{Pierpaoli} {et~al.}(2003){Pierpaoli}, {Borgani}, {Scott}, \&
  {White}}]{Pierpa03}
{Pierpaoli}, E., {Borgani}, S., {Scott}, D., \& {White}, M. 2003, \mnras, 342,
  163

\bibitem[{{Pierpaoli} {et~al.}(2001){Pierpaoli}, {Scott}, \&
  {White}}]{Pierpa01}
{Pierpaoli}, E., {Scott}, D., \& {White}, M. 2001, \mnras, 325, 77

\bibitem[{{Postman} {et~al.}(1992){Postman}, {Huchra}, \& {Geller}}]{Pos92}
{Postman}, M., {Huchra}, J.~P., \& {Geller}, M.~J. 1992, \apj, 384, 404

\bibitem[{{Postman} {et~al.}(1996){Postman}, {Lubin}, {Gunn}, {Oke}, {Hoessel},
  {Schneider}, \& {Christensen}}]{Pos96}
{Postman}, M., {Lubin}, L.~M., {Gunn}, J.~E., {Oke}, J.~B., {Hoessel}, J.~G.,
  {Schneider}, D.~P., \& {Christensen}, J.~A. 1996, \aj, 111, 615

\bibitem[{{Ramella} {et~al.}(1997){Ramella}, {Pisani}, \& {Geller}}]{Ram97}
{Ramella}, M., {Pisani}, A., \& {Geller}, M.~J. 1997, \aj, 113, 483

\bibitem[{{Romer} {et~al.}(2000){Romer}, {Nichol}, {Holden}, {Ulmer}, {Pildis},
  {Merrelli}, {Adami}, {Burke}, {Collins}, {Metevier}, {Kron}, \&
  {Commons}}]{Rom00}
{Romer}, A.~K., {Nichol}, R.~C., {Holden}, B.~P., {Ulmer}, M.~P., {Pildis},
  R.~A., {Merrelli}, A.~J., {Adami}, C., {Burke}, D.~J., {Collins}, C.~A.,
  {Metevier}, A.~J., {Kron}, R.~G., \& {Commons}, K. 2000, \apjs, 126, 209

\bibitem[{{Rosati} {et~al.}(1998){Rosati}, {della Ceca}, {Norman}, \&
  {Giacconi}}]{Ros98}
{Rosati}, P., {della Ceca}, R., {Norman}, C., \& {Giacconi}, R. 1998, \apjl,
  492, L21+

\bibitem[{{Rozo} {et~al.}(2007){Rozo}, {Wechsler}, {Koester}, {Evrard}, \&
  {McKay}}]{Roz07}
{Rozo}, E., {Wechsler}, R.~H., {Koester}, B.~P., {Evrard}, A.~E., \& {McKay},
  T.~A. 2007, ArXiv Astrophysics e-prints

\bibitem[{{Scharf} {et~al.}(2000){Scharf}, {Donahue}, {Voit}, {Rosati}, \&
  {Postman}}]{Sch00}
{Scharf}, C., {Donahue}, M., {Voit}, G.~M., {Rosati}, P., \& {Postman}, M.
  2000, \apjl, 528, L73

\bibitem[{{Schechter}(1976)}]{Sch76}
{Schechter}, P. 1976, \apj, 203, 297

\bibitem[{{Schwarz}(1978)}]{Sch78} 
{Schwarz}, U.~J.\ 1978, \aap, 65, 345 

\bibitem[{{Schneider}(1996)}]{Sch96}
{Schneider}, P. 1996, \mnras, 283, 837

\bibitem[{{Schuecker} \& {Boehringer}(1998)}]{Sch98}
{Schuecker}, P. \& {Boehringer}, H. 1998, \aap, 339, 315

\bibitem[{{Shectman}(1985)}]{She85}
{Shectman}, S.~A. 1985, \apjs, 57, 77

\bibitem[{{Sheldon} {et~al.}(2004){Sheldon}, {Johnston}, {Frieman}, {Scranton},
  {McKay}, {Connolly}, {Budav{\'a}ri}, {Zehavi}, {Bahcall}, {Brinkmann}, \&
  {Fukugita}}]{She04}
{Sheldon}, E.~S., {Johnston}, D.~E., {Frieman}, J.~A., {Scranton}, R., {McKay},
  T.~A., {Connolly}, A.~J., {Budav{\'a}ri}, T., {Zehavi}, I., {Bahcall}, N.~A.,
  {Brinkmann}, J., \& {Fukugita}, M. 2004, \aj, 127, 2544

\bibitem[{{Smith} {et~al.}(2002){Smith}, {Tucker}, {Kent}, {Richmond},
  {Fukugita}, {Ichikawa}, {Ichikawa}, {Jorgensen}, {Uomoto}, {Gunn}, {Hamabe},
  {Watanabe}, {Tolea}, {Henden}, {Annis}, {Pier}, {McKay}, {Brinkmann}, {Chen},
  {Holtzman}, {Shimasaku}, \& {York}}]{Smi02}
{Smith}, J.~A., {Tucker}, D.~L., {Kent}, S., {Richmond}, M.~W., {Fukugita}, M.,
  {Ichikawa}, T., {Ichikawa}, S.-i., {Jorgensen}, A.~M., {Uomoto}, A., {Gunn},
  J.~E., {Hamabe}, M., {Watanabe}, M., {Tolea}, A., {Henden}, A., {Annis}, J.,
  {Pier}, J.~R., {McKay}, T.~A., {Brinkmann}, J., {Chen}, B., {Holtzman}, J.,
  {Shimasaku}, K., \& {York}, D.~G. 2002, \aj, 123, 2121

\bibitem[{{Strateva} {et~al.}(2001){Strateva}, {Ivezi{\'c}}, {Knapp},
  {Narayanan}, {Strauss}, {Gunn}, {Lupton}, {Schlegel}, {Bahcall}, {Brinkmann},
  {Brunner}, {Budav{\'a}ri}, {Csabai}, {Castander}, {Doi}, {Fukugita}, {Gy{\H
  o}ry}, {Hamabe}, {Hennessy}, {Ichikawa}, {Kunszt}, {Lamb}, {McKay},
  {Okamura}, {Racusin}, {Sekiguchi}, {Schneider}, {Shimasaku}, \&
  {York}}]{Strv01}
{Strateva}, I., {Ivezi{\'c}}, {\v Z}., {Knapp}, G.~R., {Narayanan}, V.~K.,
  {Strauss}, M.~A., {Gunn}, J.~E., {Lupton}, R.~H., {Schlegel}, D., {Bahcall},
  N.~A., {Brinkmann}, J., {Brunner}, R.~J., {Budav{\'a}ri}, T., {Csabai}, I.,
  {Castander}, F.~J., {Doi}, M., {Fukugita}, M., {Gy{\H o}ry}, Z., {Hamabe},
  M., {Hennessy}, G., {Ichikawa}, T., {Kunszt}, P.~Z., {Lamb}, D.~Q., {McKay},
  T.~A., {Okamura}, S., {Racusin}, J., {Sekiguchi}, M., {Schneider}, D.~P.,
  {Shimasaku}, K., \& {York}, D. 2001, \aj, 122, 1861

\bibitem[{{Strauss} {et~al.}(2002){Strauss}, {Weinberg}, {Lupton}, {Narayanan},
  {Annis}, {Bernardi}, {Blanton}, {Burles}, {Connolly}, {Dalcanton}, {Doi},
  {Eisenstein}, {Frieman}, {Fukugita}, {Gunn}, {Ivezi{\' c}}, {Kent}, {Kim},
  {Knapp}, {Kron}, {Munn}, {Newberg}, {Nichol}, {Okamura}, {Quinn}, {Richmond},
  {Schlegel}, {Shimasaku}, {SubbaRao}, {Szalay}, {Vanden Berk}, {Vogeley},
  {Yanny}, {Yasuda}, {York}, \& {Zehavi}}]{Str02}
{Strauss}, M.~A., {Weinberg}, D.~H., {Lupton}, R.~H., {Narayanan}, V.~K.,
  {Annis}, J., {Bernardi}, M., {Blanton}, M., {Burles}, S., {Connolly}, A.~J.,
  {Dalcanton}, J., {Doi}, M., {Eisenstein}, D., {Frieman}, J.~A., {Fukugita},
  M., {Gunn}, J.~E., {Ivezi{\' c}}, {\v Z}., {Kent}, S., {Kim}, R.~S.~J.,
  {Knapp}, G.~R., {Kron}, R.~G., {Munn}, J.~A., {Newberg}, H.~J., {Nichol},
  R.~C., {Okamura}, S., {Quinn}, T.~R., {Richmond}, M.~W., {Schlegel}, D.~J.,
  {Shimasaku}, K., {SubbaRao}, M., {Szalay}, A.~S., {Vanden Berk}, D.,
  {Vogeley}, M.~S., {Yanny}, B., {Yasuda}, N., {York}, D.~G., \& {Zehavi}, I.
  2002, \aj, 124, 1810

\bibitem[{{Sutherland}(1988)}]{Sut88}
{Sutherland}, W. 1988, \mnras, 234, 159

\bibitem[{{Tremaine} \& {Richstone}(1977)}]{RT77}
{Tremaine}, S.~D. \& {Richstone}, D.~O. 1977, \apj, 212, 311

\bibitem[{{Wechsler}(2004)}]{Wec04}
{Wechsler}, R.~H. 2004, in Clusters of Galaxies: Probes of Cosmological
  Structure and Galaxy Evolution, ed. J.~S. {Mulchaey}, A.~{Dressler}, \&
  A.~{Oemler}

\bibitem[{{Weiner} {et~al.}(2005){Weiner}, {Phillips}, {Faber}, {Willmer},
  {Vogt}, {Simard}, {Gebhardt}, {Im}, {Koo}, {Sarajedini}, {Wu}, {Forbes},
  {Gronwall}, {Groth}, {Illingworth}, {Kron}, {Rhodes}, {Szalay}, \&
  {Takamiya}}]{Wei05}
{Weiner}, B.~J., {Phillips}, A.~C., {Faber}, S.~M., {Willmer}, C.~N.~A.,
  {Vogt}, N.~P., {Simard}, L., {Gebhardt}, K., {Im}, M., {Koo}, D.~C.,
  {Sarajedini}, V.~L., {Wu}, K.~L., {Forbes}, D.~A., {Gronwall}, C., {Groth},
  E.~J., {Illingworth}, G.~D., {Kron}, R.~G., {Rhodes}, J., {Szalay}, A.~S., \&
  {Takamiya}, M. 2005, \apj, 620, 595

\bibitem[{{White} \& {Kochanek}(2002)}]{Whi02}
{White}, M. \& {Kochanek}, C.~S. 2002, \apj, 574, 24

\bibitem[{{Wittman} {et~al.}(2001){Wittman}, {Tyson}, {Margoniner}, {Cohen}, \&
  {Dell'Antonio}}]{Wit01}
{Wittman}, D., {Tyson}, J.~A., {Margoniner}, V.~E., {Cohen}, J.~G., \&
  {Dell'Antonio}, I.~P. 2001, \apjl, 557, L89

\bibitem[{{Yee} {et~al.}(2000){Yee}, {Morris}, {Lin}, {Carlberg}, {Hall},
  {Sawicki}, {Patton}, {Wirth}, {Ellingson}, \& {Shepherd}}]{Yee00}
{Yee}, H.~K.~C., {Morris}, S.~L., {Lin}, H., {Carlberg}, R.~G., {Hall}, P.~B.,
  {Sawicki}, M., {Patton}, D.~R., {Wirth}, G.~D., {Ellingson}, E., \&
  {Shepherd}, C.~W. 2000, \apjs, 129, 475

\bibitem[{{York} {et~al.}(2000){York}, {Adelman}, {Anderson}, {Anderson},
  {Annis}, {Bahcall}, {Bakken}, {Barkhouser}, {Bastian}, {Berman}, {Boroski},
  {Bracker}, {Briegel}, {Briggs}, {Brinkmann}, {Brunner}, {Burles}, {Carey},
  {Carr}, {Castander}, {Chen}, {Colestock}, {Connolly}, {Crocker}, {Csabai},
  {Czarapata}, {Davis}, {Doi}, {Dombeck}, {Eisenstein}, {Ellman}, {Elms},
  {Evans}, {Fan}, {Federwitz}, {Fiscelli}, {Friedman}, {Frieman}, {Fukugita},
  {Gillespie}, {Gunn}, {Gurbani}, {de Haas}, {Haldeman}, {Harris}, {Hayes},
  {Heckman}, {Hennessy}, {Hindsley}, {Holm}, {Holmgren}, {Huang}, {Hull},
  {Husby}, {Ichikawa}, {Ichikawa}, {Ivezi{\' c}}, {Kent}, {Kim}, {Kinney},
  {Klaene}, {Kleinman}, {Kleinman}, {Knapp}, {Korienek}, {Kron}, {Kunszt},
  {Lamb}, {Lee}, {Leger}, {Limmongkol}, {Lindenmeyer}, {Long}, {Loomis},
  {Loveday}, {Lucinio}, {Lupton}, {MacKinnon}, {Mannery}, {Mantsch}, {Margon},
  {McGehee}, {McKay}, {Meiksin}, {Merelli}, {Monet}, {Munn}, {Narayanan},
  {Nash}, {Neilsen}, {Neswold}, {Newberg}, {Nichol}, {Nicinski}, {Nonino},
  {Okada}, {Okamura}, {Ostriker}, {Owen}, {Pauls}, {Peoples}, {Peterson},
  {Petravick}, {Pier}, {Pope}, {Pordes}, {Prosapio}, {Rechenmacher}, {Quinn},
  {Richards}, {Richmond}, {Rivetta}, {Rockosi}, {Ruthmansdorfer}, {Sandford},
  {Schlegel}, {Schneider}, {Sekiguchi}, {Sergey}, {Shimasaku}, {Siegmund},
  {Smee}, {Smith}, {Snedden}, {Stone}, {Stoughton}, {Strauss}, {Stubbs},
  {SubbaRao}, {Szalay}, {Szapudi}, {Szokoly}, {Thakar}, {Tremonti}, {Tucker},
  {Uomoto}, {Vanden Berk}, {Vogeley}, {Waddell}, {Wang}, {Watanabe},
  {Weinberg}, {Yanny}, \& {Yasuda}}]{Yor00}
{York}, D.~G., {Adelman}, J., {Anderson}, J.~E., {Anderson}, S.~F., {Annis},
  J., {Bahcall}, N.~A., {Bakken}, J.~A., {Barkhouser}, R., {Bastian}, S.,
  {Berman}, E., {Boroski}, W.~N., {Bracker}, S., {Briegel}, C., {Briggs},
  J.~W., {Brinkmann}, J., {Brunner}, R., {Burles}, S., {Carey}, L., {Carr},
  M.~A., {Castander}, F.~J., {Chen}, B., {Colestock}, P.~L., {Connolly}, A.~J.,
  {Crocker}, J.~H., {Csabai}, I., {Czarapata}, P.~C., {Davis}, J.~E., {Doi},
  M., {Dombeck}, T., {Eisenstein}, D., {Ellman}, N., {Elms}, B.~R., {Evans},
  M.~L., {Fan}, X., {Federwitz}, G.~R., {Fiscelli}, L., {Friedman}, S.,
  {Frieman}, J.~A., {Fukugita}, M., {Gillespie}, B., {Gunn}, J.~E., {Gurbani},
  V.~K., {de Haas}, E., {Haldeman}, M., {Harris}, F.~H., {Hayes}, J.,
  {Heckman}, T.~M., {Hennessy}, G.~S., {Hindsley}, R.~B., {Holm}, S.,
  {Holmgren}, D.~J., {Huang}, C., {Hull}, C., {Husby}, D., {Ichikawa}, S.,
  {Ichikawa}, T., {Ivezi{\' c}}, {\v Z}., {Kent}, S., {Kim}, R.~S.~J.,
  {Kinney}, E., {Klaene}, M., {Kleinman}, A.~N., {Kleinman}, S., {Knapp},
  G.~R., {Korienek}, J., {Kron}, R.~G., {Kunszt}, P.~Z., {Lamb}, D.~Q., {Lee},
  B., {Leger}, R.~F., {Limmongkol}, S., {Lindenmeyer}, C., {Long}, D.~C.,
  {Loomis}, C., {Loveday}, J., {Lucinio}, R., {Lupton}, R.~H., {MacKinnon}, B.,
  {Mannery}, E.~J., {Mantsch}, P.~M., {Margon}, B., {McGehee}, P., {McKay},
  T.~A., {Meiksin}, A., {Merelli}, A., {Monet}, D.~G., {Munn}, J.~A.,
  {Narayanan}, V.~K., {Nash}, T., {Neilsen}, E., {Neswold}, R., {Newberg},
  H.~J., {Nichol}, R.~C., {Nicinski}, T., {Nonino}, M., {Okada}, N., {Okamura},
  S., {Ostriker}, J.~P., {Owen}, R., {Pauls}, A.~G., {Peoples}, J., {Peterson},
  R.~L., {Petravick}, D., {Pier}, J.~R., {Pope}, A., {Pordes}, R., {Prosapio},
  A., {Rechenmacher}, R., {Quinn}, T.~R., {Richards}, G.~T., {Richmond}, M.~W.,
  {Rivetta}, C.~H., {Rockosi}, C.~M., {Ruthmansdorfer}, K., {Sandford}, D.,
  {Schlegel}, D.~J., {Schneider}, D.~P., {Sekiguchi}, M., {Sergey}, G.,
  {Shimasaku}, K., {Siegmund}, W.~A., {Smee}, S., {Smith}, J.~A., {Snedden},
  S., {Stone}, R., {Stoughton}, C., {Strauss}, M.~A., {Stubbs}, C., {SubbaRao},
  M., {Szalay}, A.~S., {Szapudi}, I., {Szokoly}, G.~P., {Thakar}, A.~R.,
  {Tremonti}, C., {Tucker}, D.~L., {Uomoto}, A., {Vanden Berk}, D., {Vogeley},
  M.~S., {Waddell}, P., {Wang}, S., {Watanabe}, M., {Weinberg}, D.~H., {Yanny},
  B., \& {Yasuda}, N. 2000, \aj, 120, 1579

\bibitem[{{Zehavi} {et~al.}(2004){Zehavi}, {Weinberg}, {Zheng}, {Berlind},
  {Frieman}, {Scoccimarro}, {Sheth}, {Blanton}, {Tegmark}, {Mo}, {Bahcall},
  {Brinkmann}, {Burles}, {Csabai}, {Fukugita}, {Gunn}, {Lamb}, {Loveday},
  {Lupton}, {Meiksin}, {Munn}, {Nichol}, {Schlegel}, {Schneider}, {SubbaRao},
  {Szalay}, {Uomoto}, \& {York}}]{Zeh04}
{Zehavi}, I., {Weinberg}, D.~H., {Zheng}, Z., {Berlind}, A.~A., {Frieman},
  J.~A., {Scoccimarro}, R., {Sheth}, R.~K., {Blanton}, M.~R., {Tegmark}, M.,
  {Mo}, H.~J., {Bahcall}, N.~A., {Brinkmann}, J., {Burles}, S., {Csabai}, I.,
  {Fukugita}, M., {Gunn}, J.~E., {Lamb}, D.~Q., {Loveday}, J., {Lupton}, R.~H.,
  {Meiksin}, A., {Munn}, J.~A., {Nichol}, R.~C., {Schlegel}, D., {Schneider},
  D.~P., {SubbaRao}, M., {Szalay}, A.~S., {Uomoto}, A., \& {York}, D.~G. 2004,
  \apj, 608, 16

\bibitem[{{Zheng} {et~al.}(2005)}]{Zhe05} {Zheng}, Z., et al.\ 2005, 
\apj, 633, 791 

\bibitem[{{Zwicky} {et~al.}(1968){Zwicky}, {Herzog}, \& {Wild}}]{Zwi68}
{Zwicky}, F., {Herzog}, E., \& {Wild}, P. 1968, {Catalogue of galaxies and of
  clusters of galaxies} (Pasadena: California Institute of Technology (CIT),
  1961-1968)

\end{thebibliography}
\bibliographystyle{apj}

\clearpage


\epsscale{1.00}
\begin{figure}
\plottwo{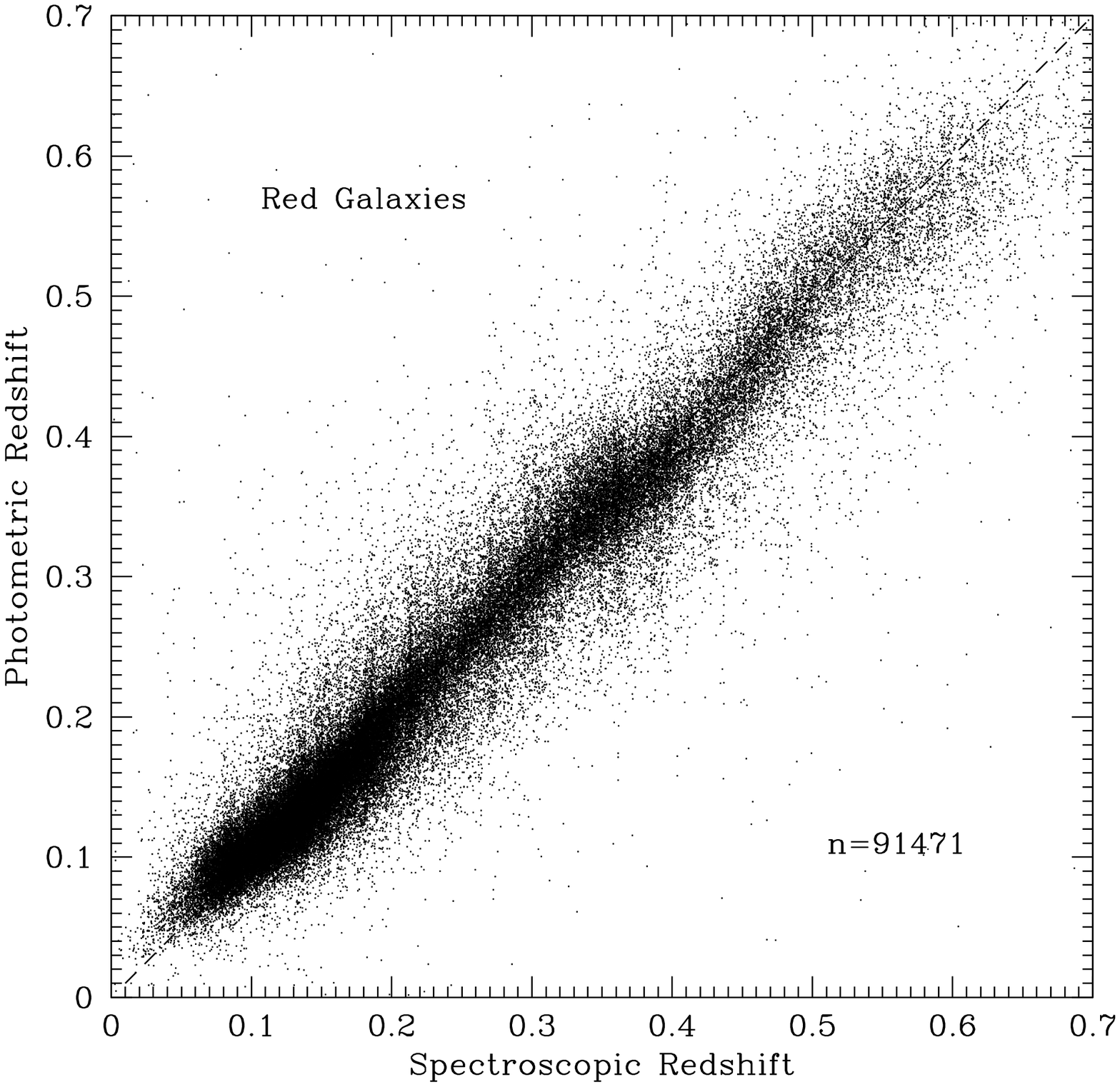}{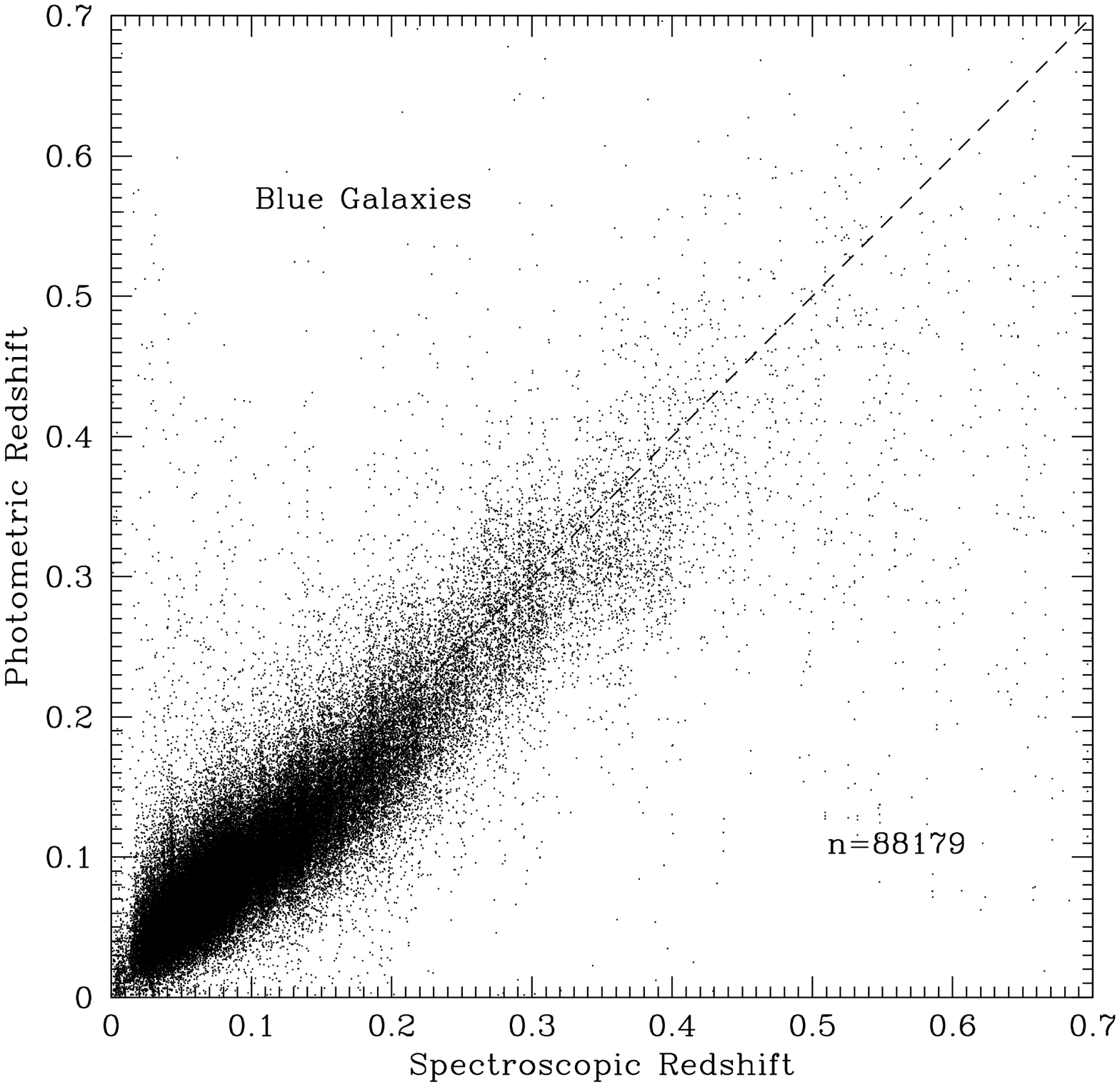}
\caption{Calculated photometric redshifts versus corresponding spectroscopic measurements 
for early type galaxies (or red galaxies, left), and late type galaxies (or blue galaxies, 
right). Here, red means $g-r >$ 1.3 and blue means $g-r <$ 1.3.
\label{f1}}
\end{figure}

\epsscale{0.75}
\begin{figure}
\plottwo{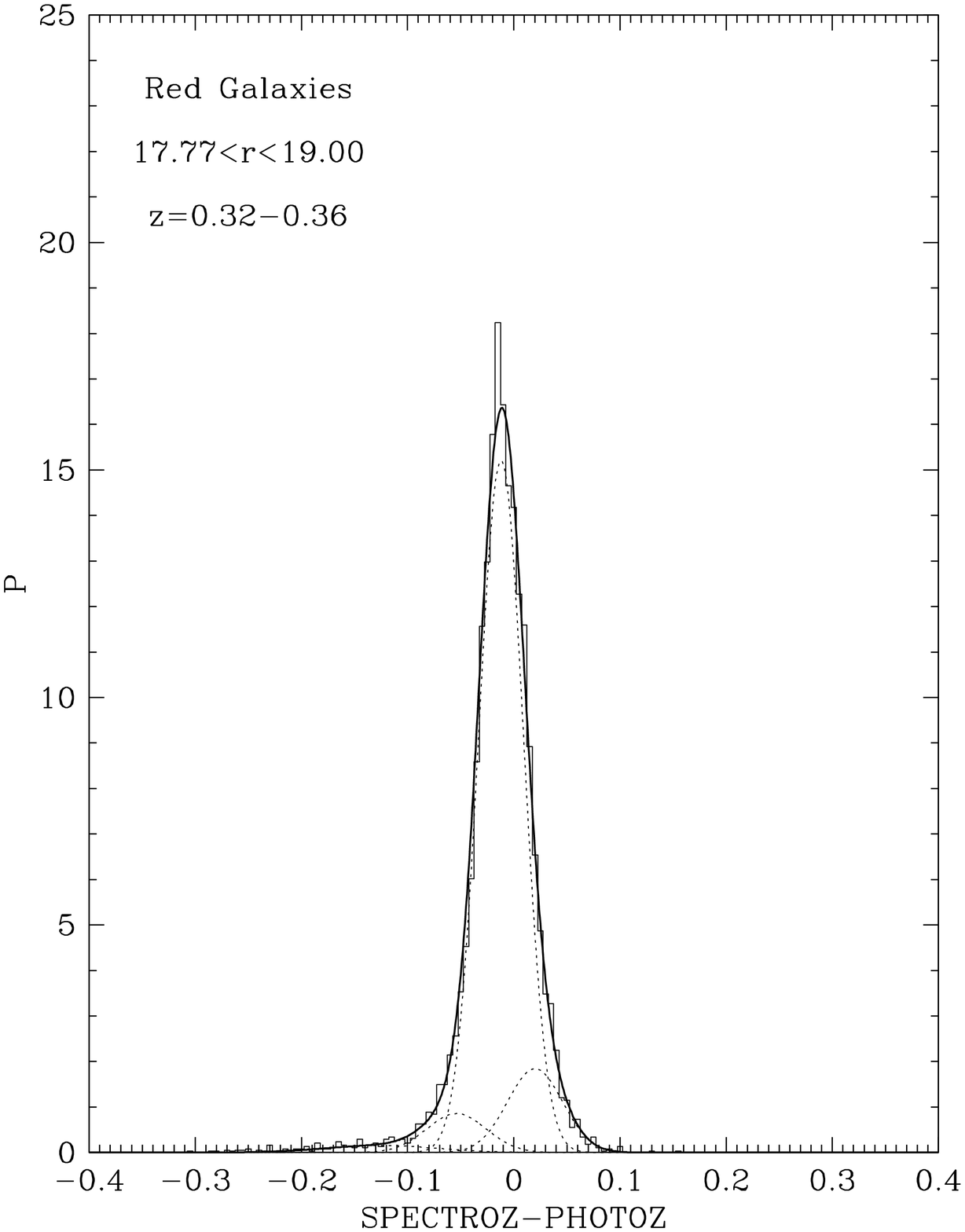}{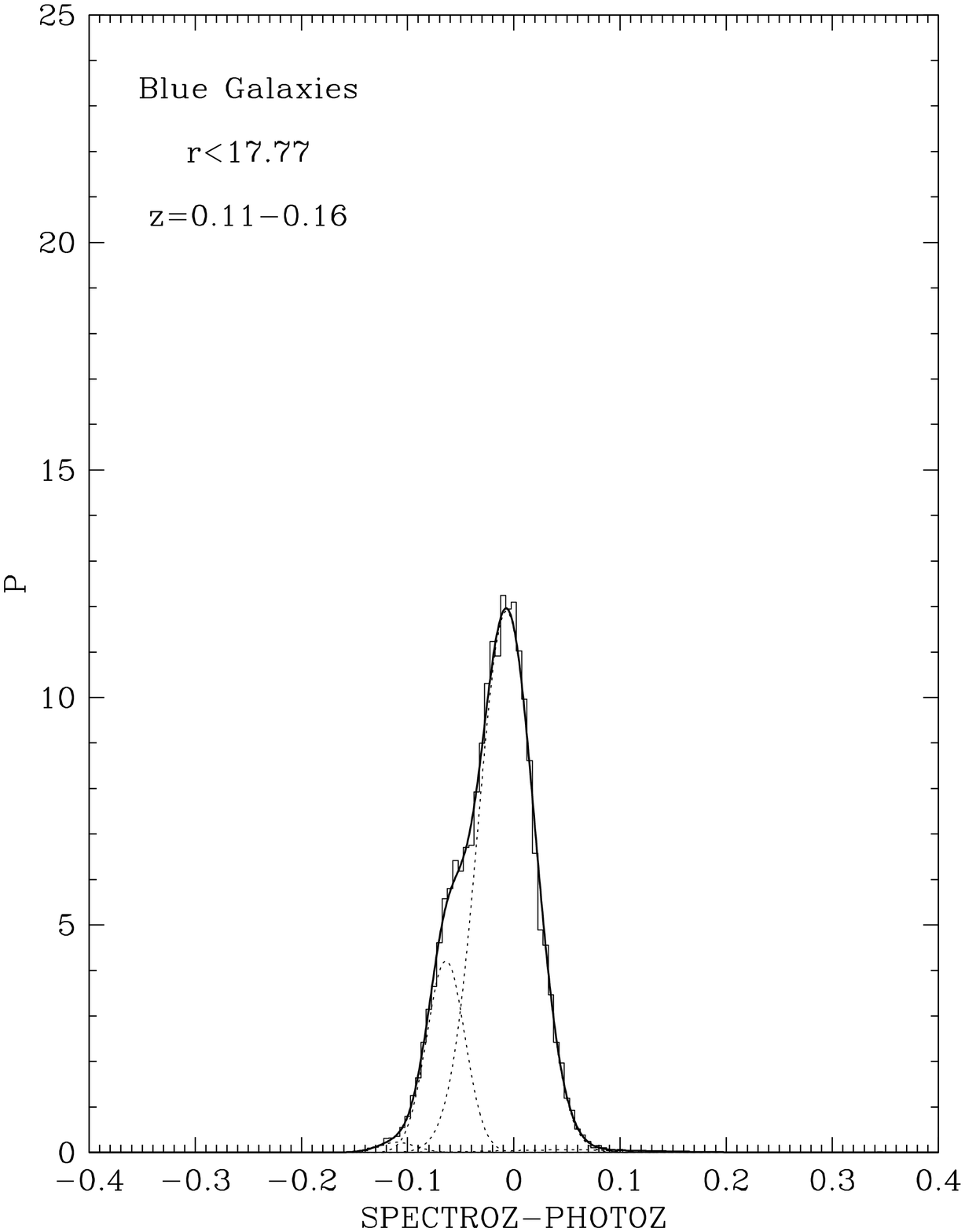}
\caption{Examples of multiple Gaussian fits for the error distributions of computed 
photometric redshifts. The derived fitting parameters are used to scatter the known 
redshifts of mock galaxies in order to simulate the practice with real SDSS data. 
\label{f2}}
\end{figure}

\epsscale{1.0}
\begin{figure}
\plottwo{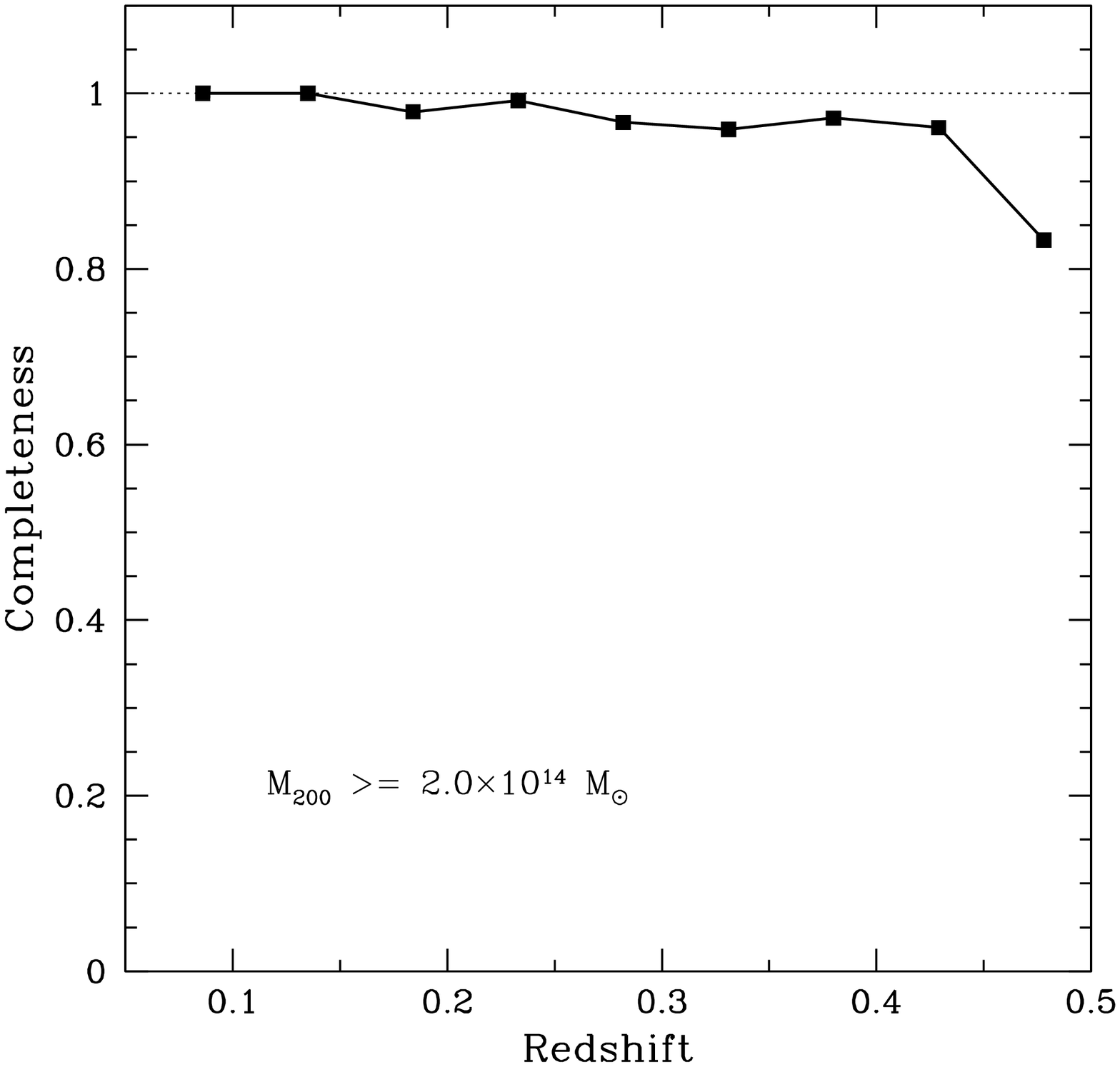}{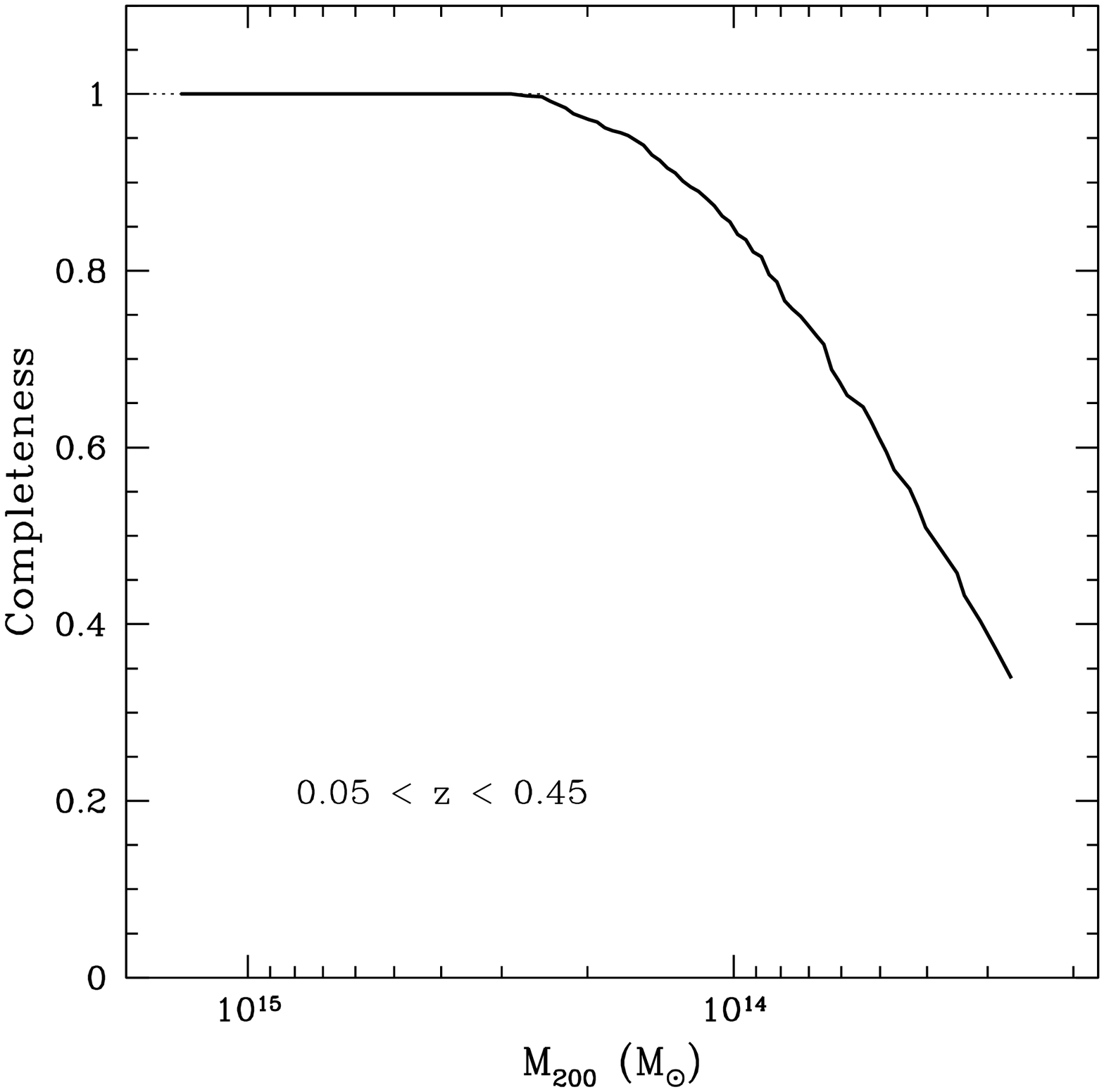}
\caption{Completeness of the detected cluster sample as a function of redshift 
and the virial mass of matched halos, respectively. The sample shows 
a consistent completeness of $>95\%$ complete for halos with 
$M_{200} > 2.0\times10^{14} h^{-1} M_\odot$ and is $\sim85\%$ complete for 
halos with $M_{200} > 1.0\times10^{14} h^{-1} M_\odot$ in the redshift range of 
$0.05<z<0.45$. Note that the annotations in the figures should read $h^{-1} M_\odot$ 
instead of $M_\odot$.
\label{f3}}
\end{figure}

\epsscale{1.0}
\begin{figure}
\plottwo{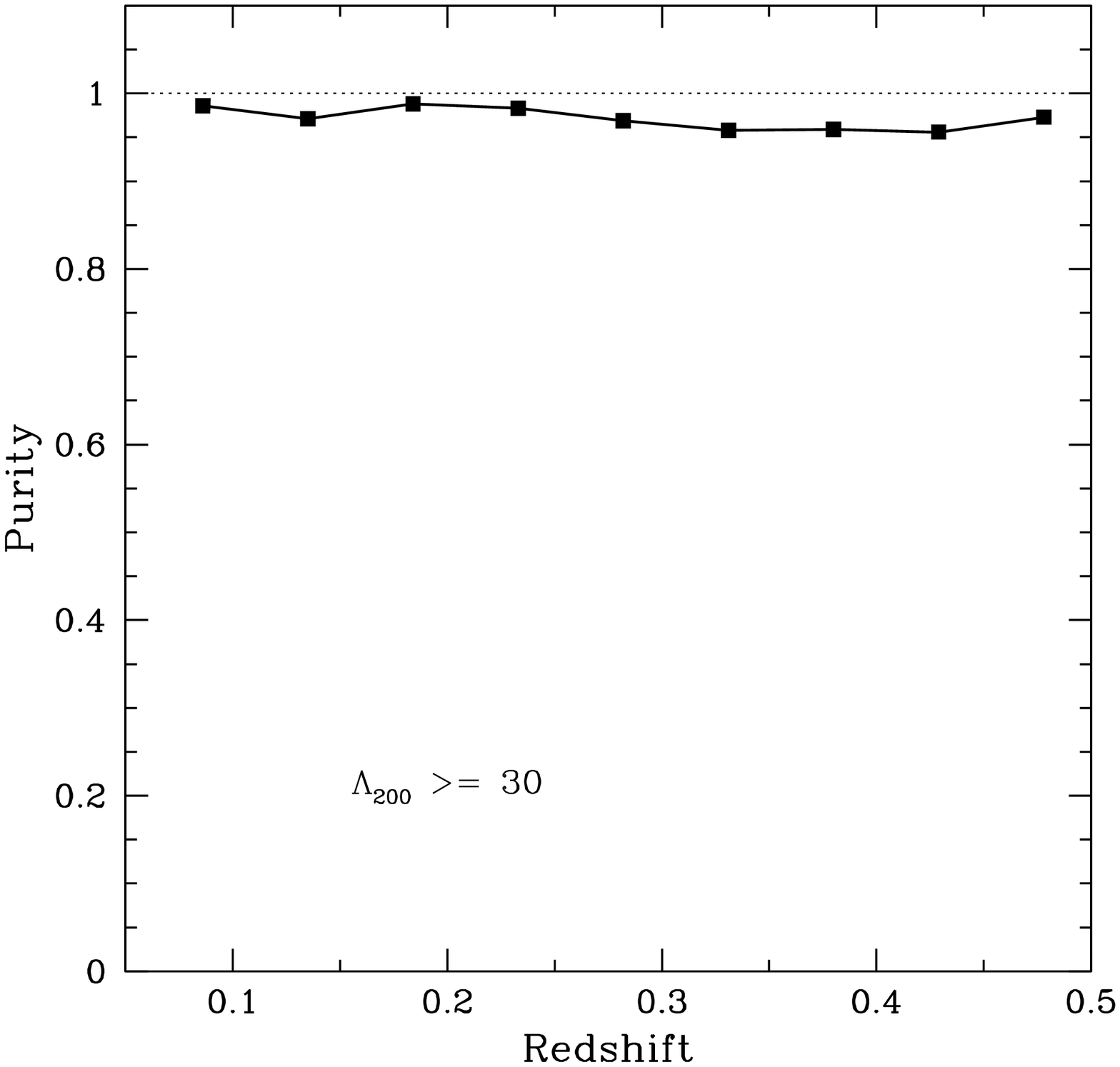}{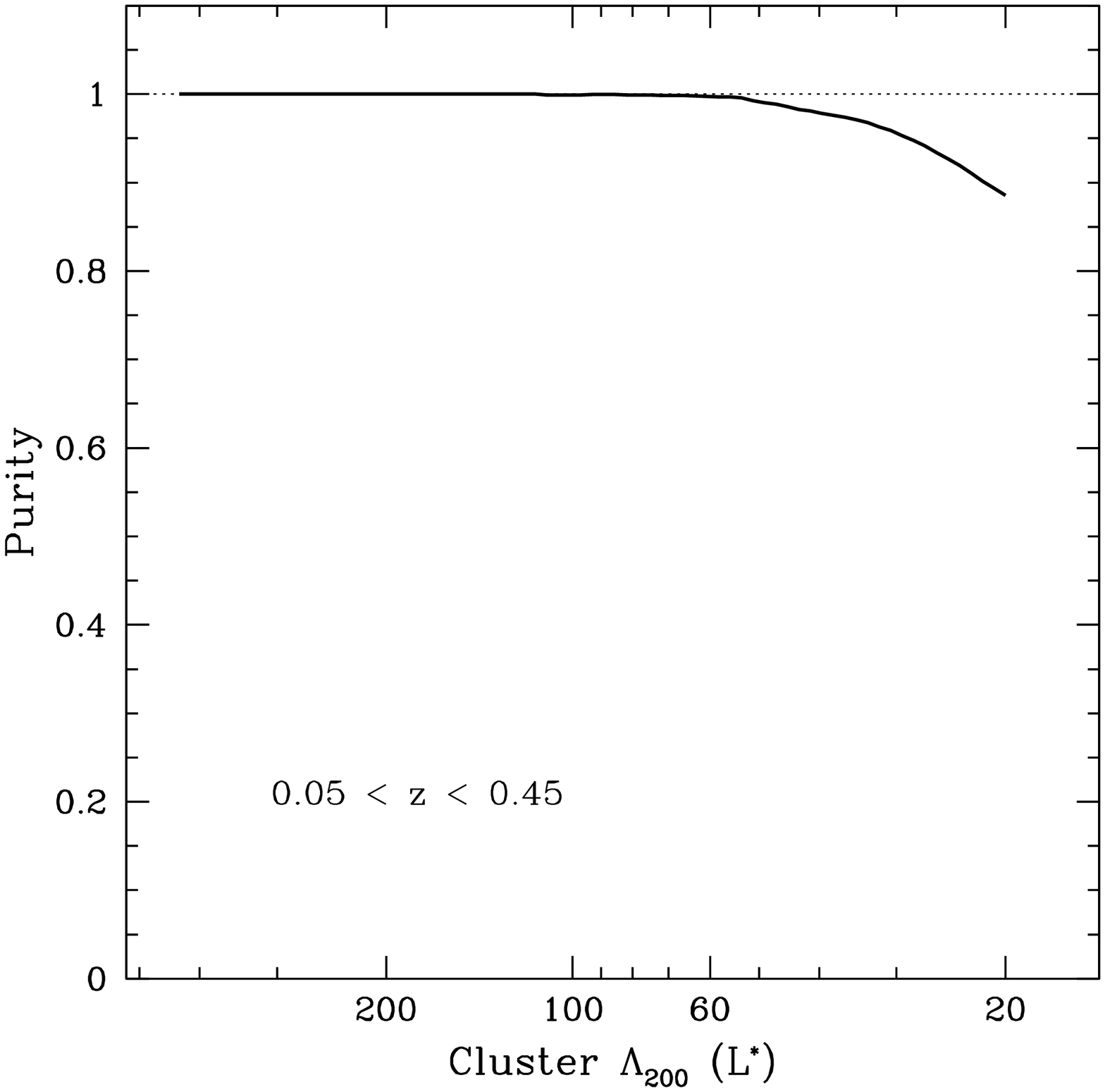}
\caption{Purity of the detected cluster sample as a function of redshift 
and the cluster richness, respectively. The derived catalog is over $95\%$ pure for 
clusters with $\Lambda_{200} > 30$ and around $90\%$ pure for $\Lambda_{200} > 20$ 
in the redshift range of $0.05<z<0.45$.
\label{f4}}
\end{figure}

\epsscale{1.00}
\begin{figure}
\plotone{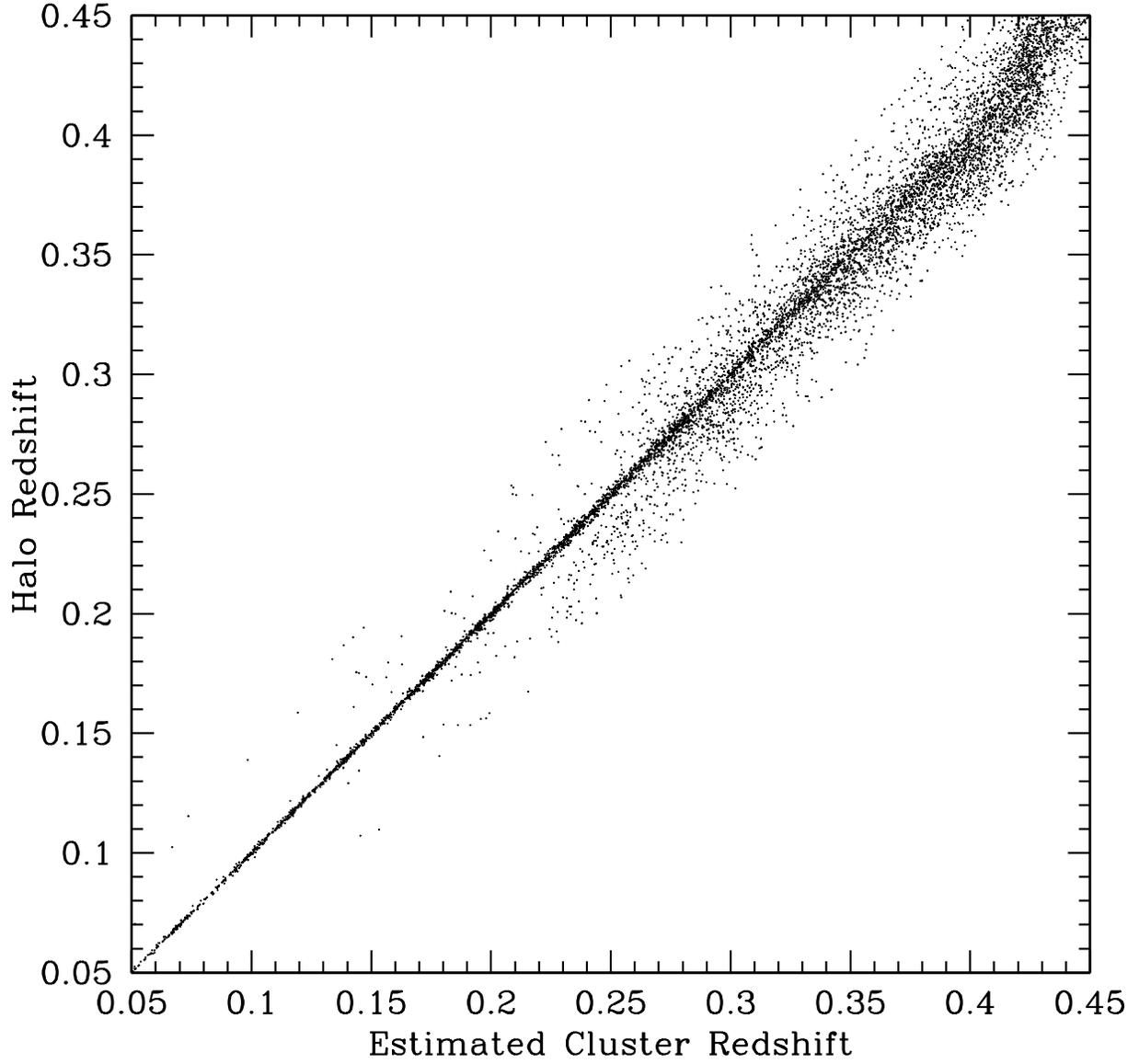}
\caption{Comparison between estimated cluster redshifts and known redshifts of 
matched halos.
\label{f5}}
\end{figure}

\epsscale{0.87}
\begin{figure}
\plotone{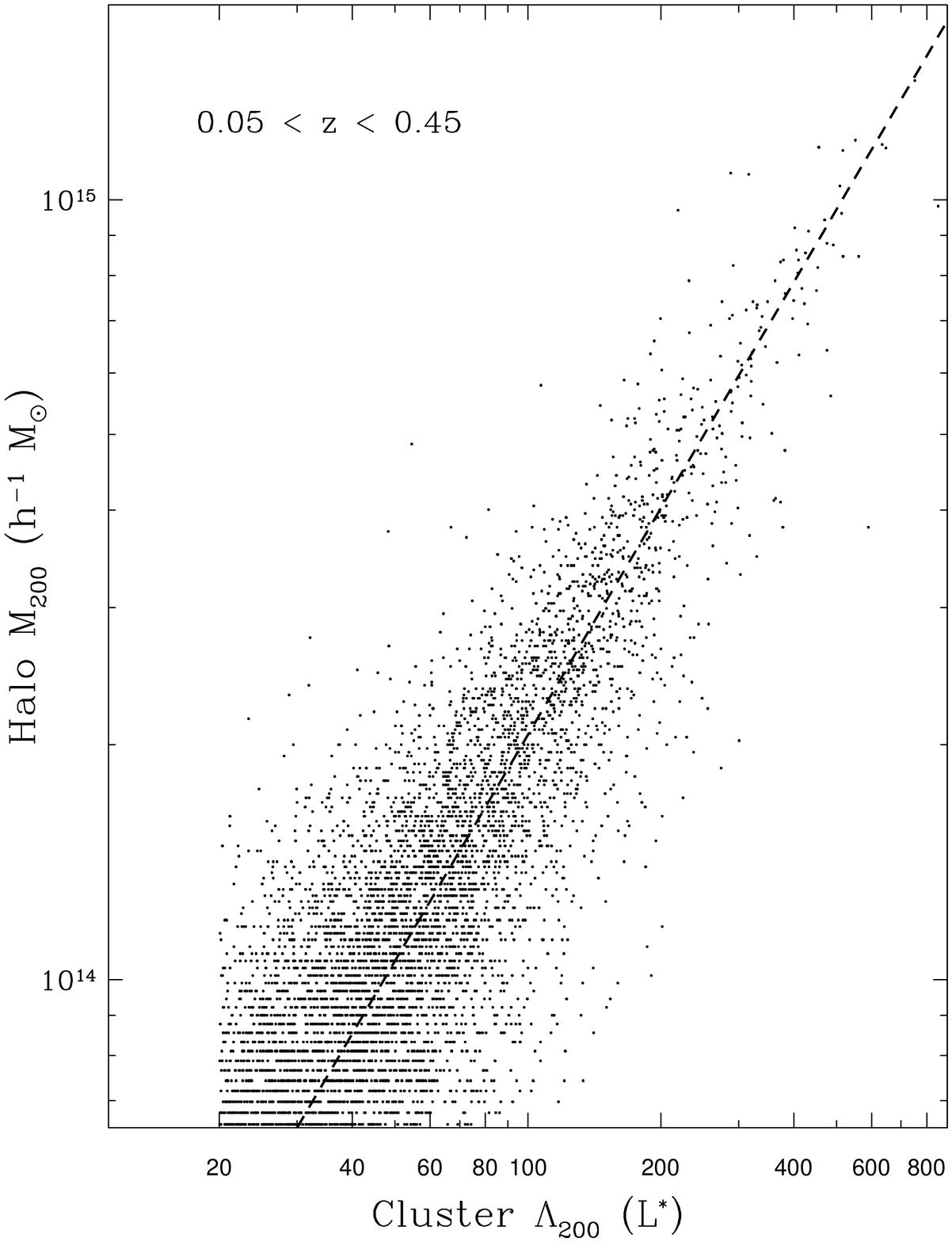}
\caption{Comparison between derived cluster richness and the virial mass of matched 
halos. The cluster richness $\Lambda_{200}$ is the total luminosity 
of the cluster in units of $L^*$ inside its virial radius $r_{200}$.  
\label{f6}}
\end{figure}

\epsscale{0.87}
\begin{figure}
\plotone{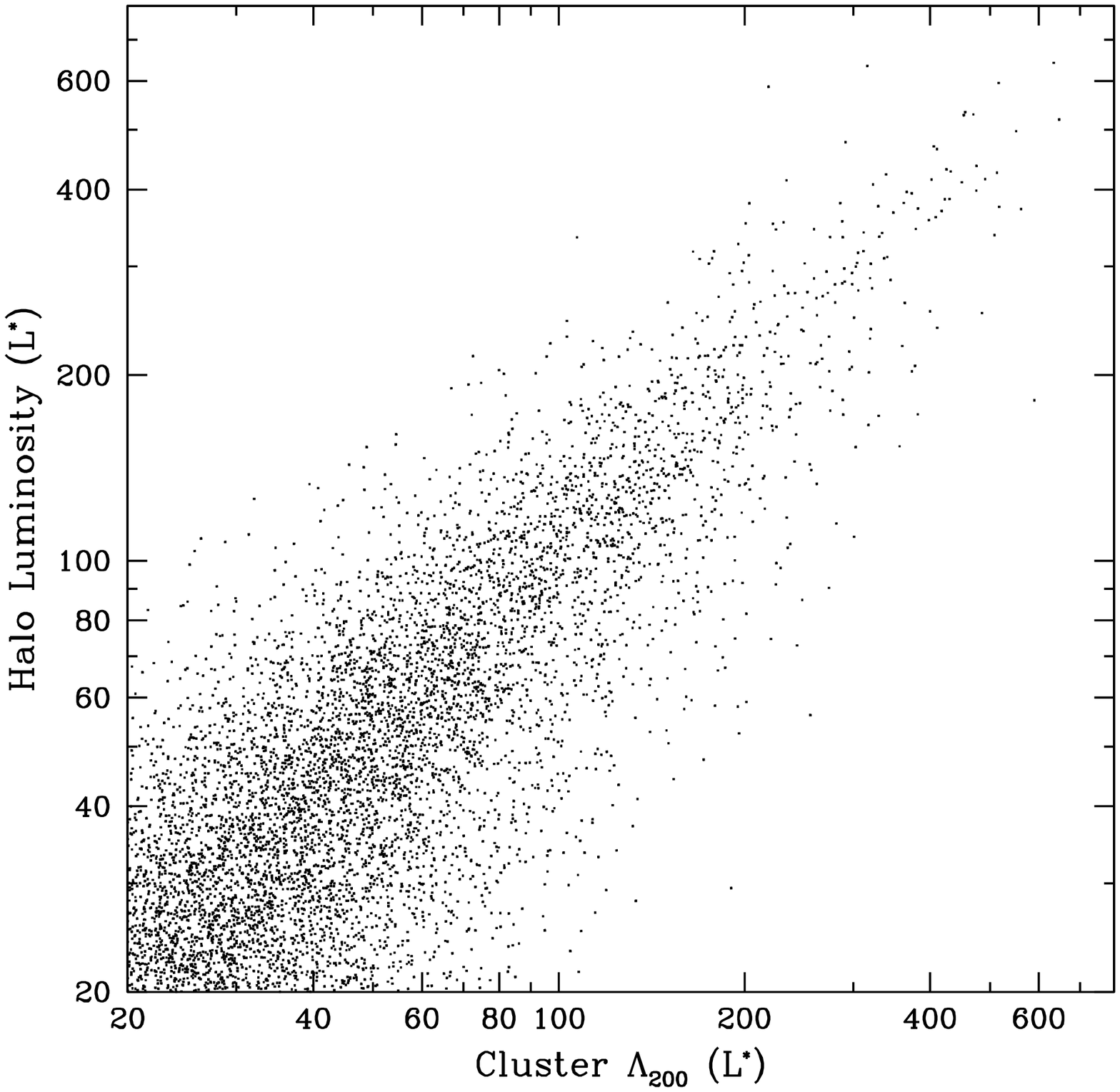}
\caption{Comparison between derived cluster richness and the total luminosity of matched 
halos in units of $L^*$. The cluster richness $\Lambda_{200}$ is the total luminosity 
of the cluster in units of $L^*$ inside its virial radius $r_{200}$.  
\label{f7}}
\end{figure}

\epsscale{0.87}
\begin{figure}
\plotone{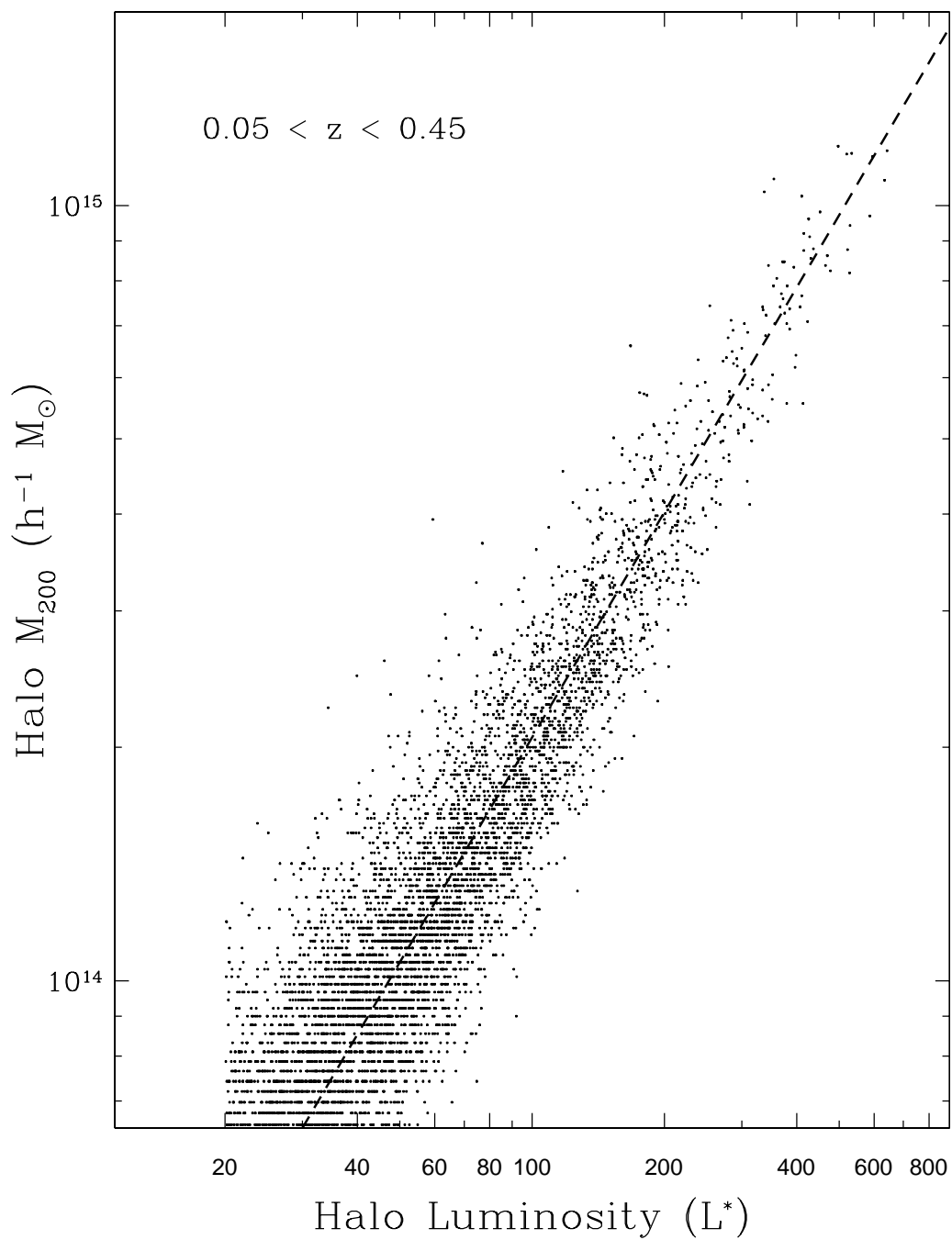}
\caption{Comparison between the virial mass of matched halos and their luminosities in 
units of $L^*$. The dashed line is the best-fit cluster richness-mass scaling relation 
given in Figure 6. 
\label{f8}}
\end{figure}

\epsscale{0.87}
\begin{figure}
\plotone{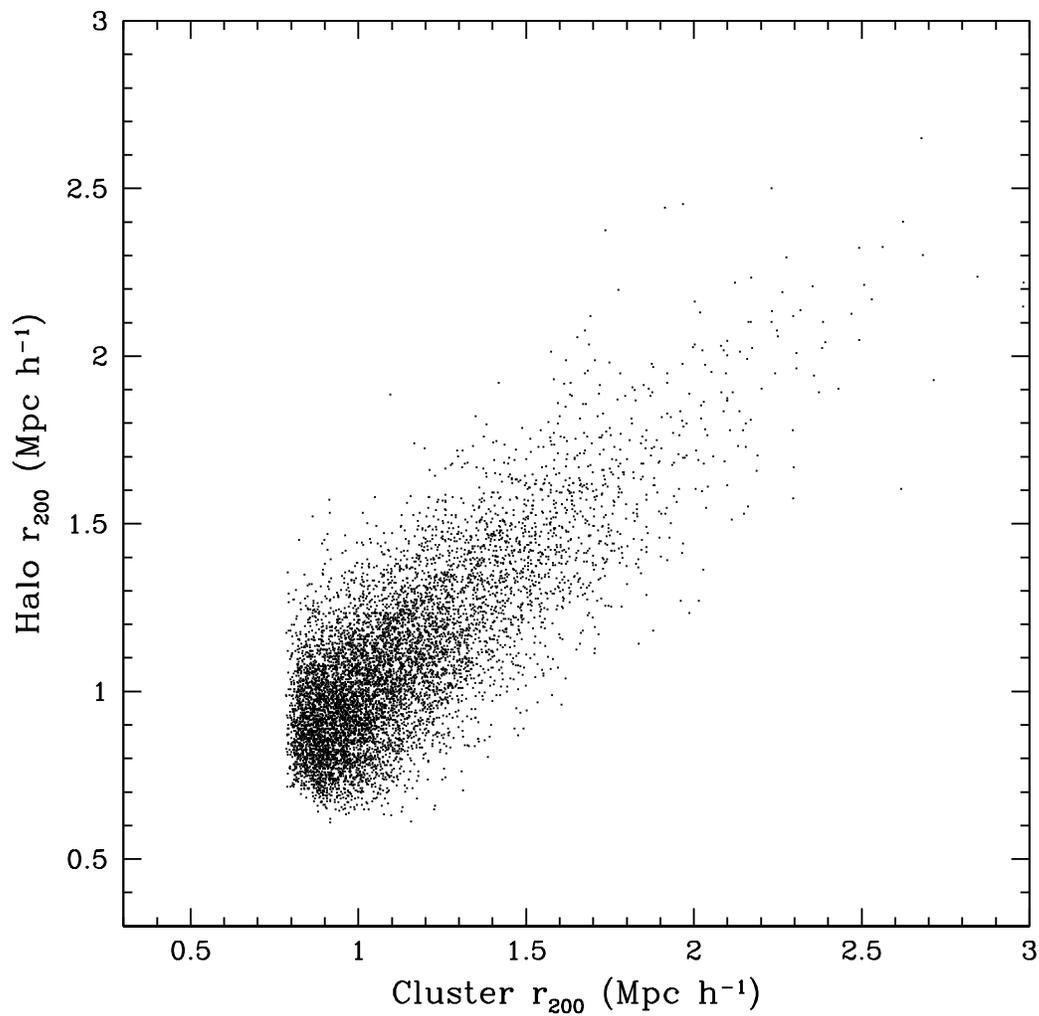}
\caption{Comparison between derived cluster virial radius $r_{200}$ and the halo 
$r_{200}$ determined by galaxy overdensities.
\label{f9}}
\end{figure}

\end{document}